\documentclass{ctr}

\usepackage{ctrfont}
\usepackage{natbib}

\usepackage{url}
\usepackage{amsmath}
\usepackage{amssymb}

\usepackage{graphicx}

\usepackage{subfig}

\usepackage[labelsep=period]{caption}
\renewcommand{\tablename}{\sc Table}
\renewcommand{\figurename}{\sc Figure}

\title{Performance of wall-modeled LES for external aerodynamics in the NASA Juncture Flow}

\shorttitle{Performance of WMLES for external aerodynamics}

\author{A. Lozano-Dur\'an, S.~T. Bose \and P. Moin}

\shortauthor{Lozano-Dur\'an, Bose \& Moin}

\begin{document}


\maketitle


\section{Motivation and objectives}

The use of computational fluid dynamics (CFD) for external aerodynamic
applications has been a key tool for aircraft design in the modern
aerospace industry. CFD methodologies with increasing functionality
and performance have greatly improved our understanding and predictive
capabilities of complex flows.  These improvements suggest that
Certification by Analysis (CbA) --prediction of the aerodynamic
quantities of interest by numerical simulations~\citep{Clark2020} may
soon be a reality.  CbA is expected to narrow the number of wind
tunnel experiments, reducing both the turnover time and cost of the design
cycle. However, flow predictions from the state-of-the-art CFD solvers
are still unable to comply with the stringent accuracy requirements
and computational efficiency demanded by the industry. These
limitations are imposed, largely, by the defiant ubiquity of
turbulence.  In the present work, we investigate the performance of
wall-modeled large-eddy simulation (WMLES) to predict the mean flow
quantities over the fuselage and wing-body junction of the NASA
Juncture Flow Experiment \citep{Rumsey2019}.


Computations submitted to previous AIAA Drag Prediction
Workshops~\citep{Vassberg2008} have displayed large variations in the
prediction of separation, skin friction, and pressure in the
corner-flow region near the wing trailing edge. To improve the
performance of CFD, NASA has developed a validation experiment for a
generic full-span wing-fuselage junction model at subsonic
conditions. The reader is referred to \citet{Rumsey2019} for a summary
of the history and goals of the NASA Juncture Flow
Experiment~\citep[see also][]{Rumsey2016a, Rumsey2016b}. The geometry
and flow conditions are designed to yield flow separation in the
trailing edge corner of the wing, with recirculation bubbles varying
in size with the angle of attack (AoA). The model is a full-span
wing-fuselage body that was configured with truncated DLR-F6 wings,
both with and without leading-edge horn at the wing root. The model
has been tested at a chord Reynolds number of 2.4 million, and
AoA ranging from -10 degrees to +10 degrees in the Langley 14- by
22-foot Subsonic Tunnel. An overview of the experimental measurements
can be found in \citet{Kegerise2019}. The main aspects of the planning
and execution of the project are discussed by \citet{Rumsey2018},
along with details about the CFD and experimental teams.

To date, most CFD efforts on the NASA Juncture Flow Experiment have
been conducted using RANS or hybrid-RANS solvers.  \citet{Lee2017}
performed the first CFD analysis to aid the NASA Juncture Flow
committee in selecting the wing configuration for the final
experiment. \citet{Lee2018} presented a preliminary CFD study of the
near wing-body juncture region to evaluate the best practices in
simulating wind tunnel effects.  \citet{Rumsey2019} used FUN3D to
investigate the ability of RANS-based CFD solvers to predict the flow
details leading up to separation. The study comprised different RANS
turbulence models such as a linear eddy viscosity one-equation model,
a nonlinear version of the same model, and a full second-moment
seven-equation model. \citet{Rumsey2019} also performed a grid
sensitivity analysis and CFD uncertainty quantification. Comparisons
between CFD simulations and the wind tunnel experimental results have
been recently documented by \citet{Lee2019}.

NASA has recognized WMLES as a critical pacing item for ``developing a
visionary CFD capability required by the notional year
2030''. According to NASA's recent CFD Vision 2030 report
\citep{Slotnick2014}, hybrid RANS/LES \citep{Spalart1997, Spalart2009}
and WMLES \citep{Bose2018} are identified as the most viable
approaches for predicting realistic flows at high Reynolds numbers in
external aerodynamics. However, WMLES has been less thoroughly
investigated. In the present study, we perform WMLES of the NASA
Juncture Flow. Other attempts of WMLES of the same flow configuration
include the works by \cite{Iyer2020}, \cite{Ghate2020}, and
\cite{Lozano_AIAA_2020}.  These authors highlighted the capabilities
of WMLES for predicting wall pressure, velocity and Reynolds stresses,
especially compared with RANS-based methodologies. Nonetheless, it was
noted that WMLES is still far from providing the robustness and
stringent accuracy required for CbA, especially in the separated
regions and wing-fuselage corners. The goal of this brief is to
systematically quantify some of these errors. Modeling improvements to
alleviate these limitations are discussed in the companion brief by
\cite{Lozano_brief_2020_2}, which can be found in this volume.


This brief is organized as follows. The flow setup, mathematical
modeling, and numerical approach are presented in Section
\ref{sec:numerical}. The strategies for grid generation are discussed
in Section \ref{sec:gridding}. The results are presented in Section
\ref{sec:results}, which includes the prediction and error scaling of
the mean velocity profiles and Reynolds stresses for three different
locations on the aircraft: the upstream region of the fuselage, the
wing-body juncture, and the wing-body juncture close to the
trailing edge. Finally, conclusions are offered in Section
\ref{sec:conclusions}.

\section{Numerical Methods}\label{sec:numerical}

\subsection{Flow conditions and computational setup}\label{sec:setup}

We use the NASA Juncture Flow geometry with a wing based on the DLR-F6
and a leading-edge horn to mitigate the effect of the horseshoe vortex
over the wing-fuselage juncture.  The model wingspan is nominally
3397.2~mm, the fuselage length is 4839.2~mm, and the crank chord
(chord length at the Yehudi break) is $L=557.1$~mm. The frame of
reference is such that the fuselage nose is located at $x = 0$, the
$x$-axis is aligned with the fuselage centerline, the $y$-axis denotes
spanwise direction, and the $z$-axis is the vertical direction.  The
associated instantaneous velocities are denoted by $u$, $v$, and $w$,
and occasionally by $u_1$, $u_2$, and $u_3$.

In the experiment, the model was tripped near the front of the
fuselage and on the upper and lower surfaces of both wings.  In our
case, preliminary calculations showed that tripping was also necessary
to trigger the transition to turbulence over the wing. Hence, the
geometry of the wing was modified by displacing in the $z$ direction a
line of surface mesh points close to the leading edge by 1~mm along the
suction side of the wing, and by -1~mm along the pressure side. The
tripping lines follow approximately the location of the tripping dots
used in the experimental setup for the left wing (lower surface $x =
(4144-y)/2.082$; upper surface $x = (3775-y)/1.975$ for $y<-362$ and
$x = (2847-y)/1.532$ for $y>-362$). Tripping using dots mimicking the
experimental setup was also tested.  It was found that the results
over the wing-body juncture show little sensitivity to the tripping
due to the presence of the incoming boundary layer from the
fuselage. No tripping was needed on the fuselage, which naturally
transitioned from laminar to turbulence.

In the wind tunnel, the model was mounted on a sting aligned with the
fuselage axis.  The sting was attached to a mast that emerged from the
wind tunnel floor. Here, all calculations are performed in free air
conditions, and the sting and mast are ignored. The computational
setup is such that the dimensions of the domain are about five times the
length of the fuselage  in the three directions. The Reynolds number is
$Re = L U_\infty/\nu=2.4$ million based on the crank chord length $L$,
and freestream velocity $U_\infty$. The freestream Mach number is Ma
$= 0.189$, the freestream temperature is $T = 288.84$ K, and the dynamic
pressure is 2476 Pa.  We impose a uniform plug flow as the inflow
boundary condition. The Navier--Stokes characteristic boundary
condition for subsonic non-reflecting outflow is imposed at the
lateral boundaries, outflow and top boundaries~\citep{Poinsot1992}.  At
the walls, we impose Neumann boundary conditions with the shear stress
provided by the wall model as described in Section \ref{sec:models}.
%

\subsection{Subgrid-scale and wall modeling}\label{sec:models}

The simulations are conducted with the high-fidelity solver charLES
developed by Cascade Technologies, Inc. The code integrates the
compressible LES equations using a kinetic-energy conserving,
second-order accurate, finite volume method. The numerical
discretization relies on a flux formulation which is approximately
entropy preserving in the inviscid limit, thereby limiting the amount
of numerical dissipation added into the calculation. The time
integration is performed with a third-order Runge-Kutta explicit
method. The SGS model is the dynamic Smagorinsky model
\citep{Germano1991} with the modification by \cite{Lilly1992}.

We utilize a wall model to overcome the restrictive grid-resolution
requirements to resolve the small-scale flow motions in the vicinity
of the walls. The no-slip boundary condition at the walls is replaced
by a wall-stress boundary condition.  The wall stress is obtained from
the wall model and the walls are assumed isothermal. We use an
algebraic equilibrium wall model derived from the integration of the
one-dimensional equilibrium stress model along the wall-normal
direction \citep{Wang2002, Kawai2012, Larsson2015}. The matching
location for the wall model is the first off-wall cell center of the
LES grid. No temporal filtering or treatments were used.

\section{Grid strategies and cost}\label{sec:gridding}

\subsection{Grid generation: constant-size grid vs. boundary-layer-conforming grid}
\label{subsec:tbl}

The mesh generation is based on a Voronoi hexagonal close-packed
point-seeding method. We follow two strategies for grid generation:
\begin{itemize}
\item[(i)] Constant-size grid. In the first
  approach, we set the grid size in the vicinity of the aircraft
  surface to be roughly isotropic $\Delta\approx \Delta_x \approx
  \Delta_y \approx \Delta_z$. The number of layers in the direction
  normal to the wall of size $\Delta$ is also specified. We set the
  far-field grid resolution, $\Delta_\mathrm{far}>\Delta$, and create
  additional layers with increasing grid size to blend the near-wall
  grid with the far-field size. The meshes are constructed using a
  Voronoi diagram and ten iterations of Lloyd's algorithm to smooth
  the transition between layers with different grid
  resolutions. Figure \ref{fig:tbl_grids}(a) illustrates the grid
  structure for $\Delta=2$~mm and $\Delta_\mathrm{far}=80$~mm.  This
  grid-generation approach is algorithmically simple and
  efficient. However, it is agnostic to details of the actual flow
  such as wake/shear regions and boundary-layer growth. This implies
  that flow regions close to the fuselage nose and wing leading edge
  are underresolved (less than one point per boundary-layer
  thickness), whereas the wing trailing edge and the downstream-fuselage
  regions are seeded with up to hundreds of points per boundary-layer
  thickness. The gridding strategy (ii) aims at alleviating this issue.
%
\begin{figure}
  \begin{center}
    \includegraphics[width=1.0\textwidth]{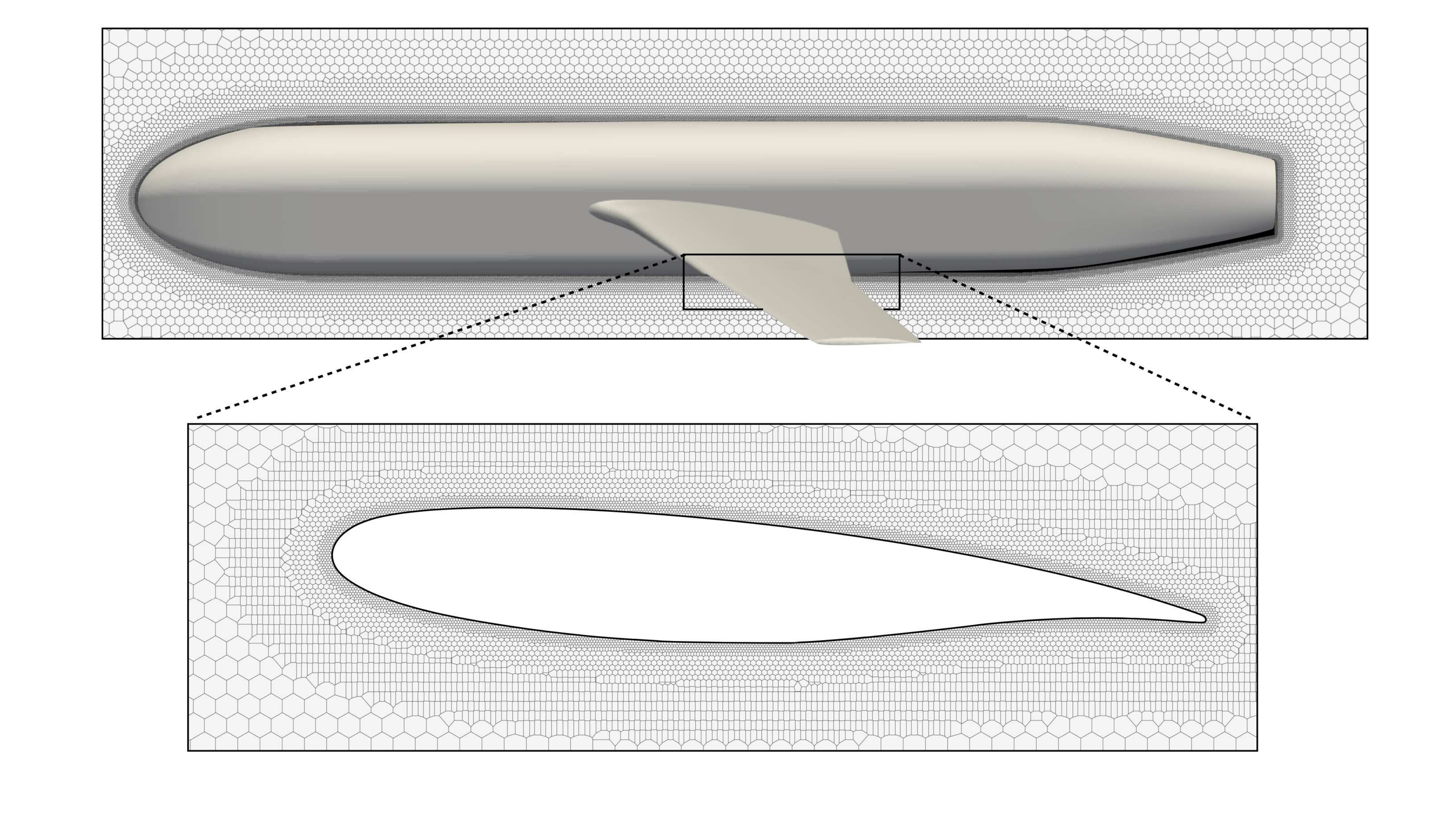}
  \end{center}
    \begin{center}
  \includegraphics[width=1.0\textwidth]{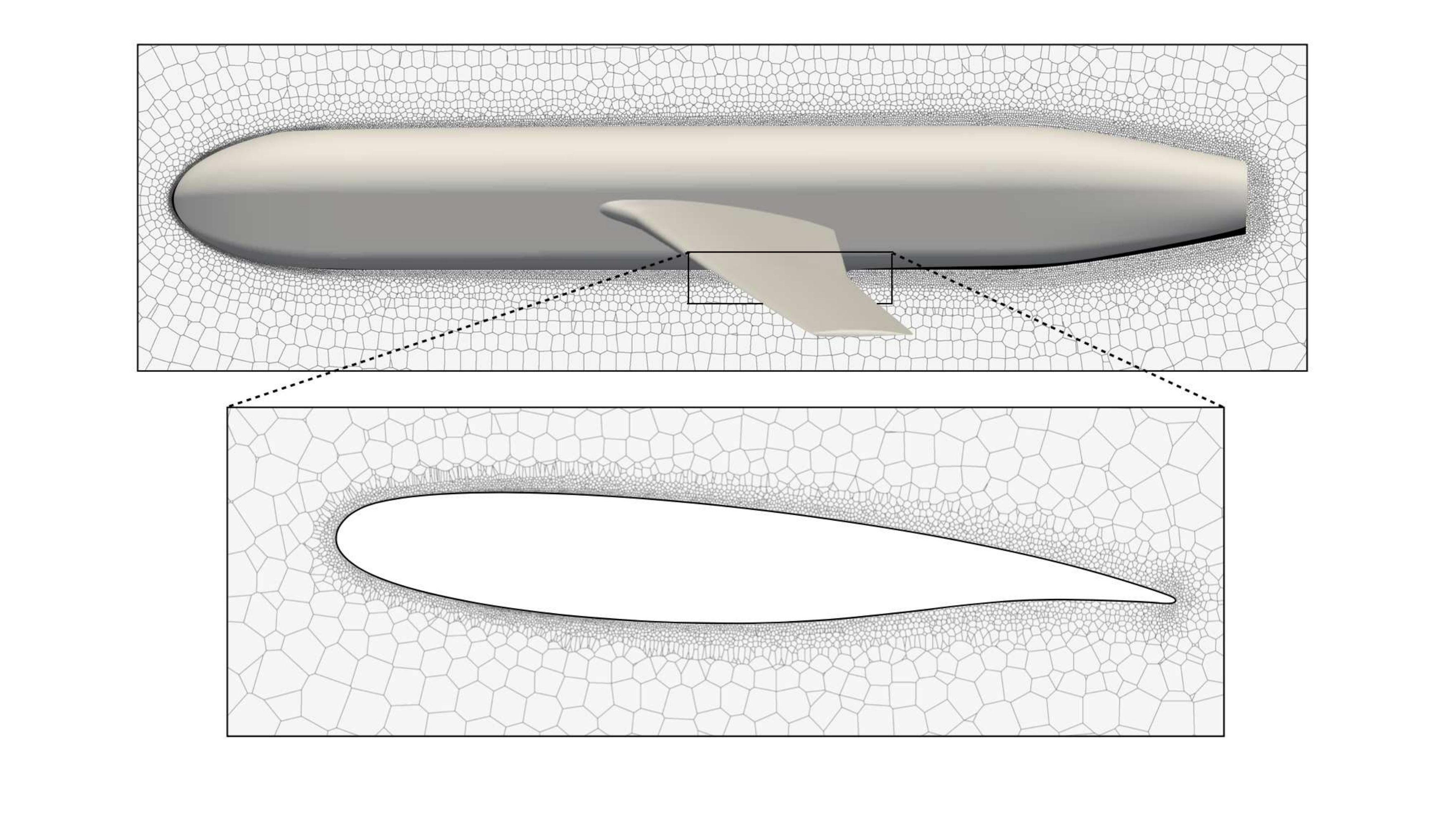}
  \end{center}
  \caption{Visualization of Voronoi control volumes for (top)
    constant-size grid following strategy i) with $\Delta=2$~mm and
    $\Delta_\mathrm{far}=80$~mm and (bottom) boundary-layer-conforming grid
    following strategy ii) with $N_{bl} = 5$ and
    $Re_\Delta^\mathrm{min}=8 \times 10^3$.\label{fig:tbl_grids}}
\end{figure}
%
\item[(ii)] Boundary-layer-conforming grid. In the second
  gridding strategy, we account for the actual growth of the turbulent
  boundary layers $\delta$ by seeding the control volumes
  consistently. We refer to this approach as boundary-layer-conforming
  grid (BL-conforming grid). The method necessitates two
  parameters. The first one is the number of points per boundary-layer
  thickness, $N_{bl}$, such that $\Delta_x \approx \Delta_y \approx
  \Delta_z \approx \Delta \approx \delta/N_{bl}$, which is a function
  of space.  The second parameter is less often discussed in the
  literature and is the minimum local Reynolds number that we are
  willing to represent in the flow, $Re_{\Delta}^\mathrm{min} \equiv
  \Delta_\mathrm{min} U_\infty/\nu$, where $\Delta_\mathrm{min}$ is
  the smallest grid resolution allowed. This is a necessary constraint
  as $\delta \rightarrow 0$ at the body leading edge, which would
  impose a large burden on the number of points required.  The last
  condition is a geometric constraint such that $\Delta$ is smaller
  than the local radius of curvature $R$ of the surface. The grid is
  then constructed by seeding points within the boundary layer with
  space-varying grid size
  \begin{equation}
    \Delta(x,y,z) \approx \mathrm{min}\left[ \mathrm{max}\left( 
    \frac{\gamma\delta}{N_{tbl}}, \frac{Re_{\Delta}^\mathrm{min} \nu}{U_\infty}\right), \beta R \right],
  \end{equation}
  where $\gamma=1.2$ is a correction factor for $\delta$ to ensure the
  near-wall grid contains the instantaneous boundary layer, and
  $\beta=1/2$. Note that the grid is still locally isotropic and the
  characteristic size of the control volumes is $\delta/N_{tbl}$ in
  the three spatial directions. Figure \ref{fig:tbl_grids}(b) shows
  the structure of a BL-conforming grid with $N_{bl} = 5$
  and $Re_\Delta^\mathrm{min}=8 \times 10^3$. Additional control
  volumes of increasing size are created to blend the near-wall grid
  with the far-field grid of size $\Delta_{\mathrm{far}}$.
\end{itemize}

\begin{figure}
\begin{center}
  \subfloat[]{\includegraphics[width=0.47\textwidth]{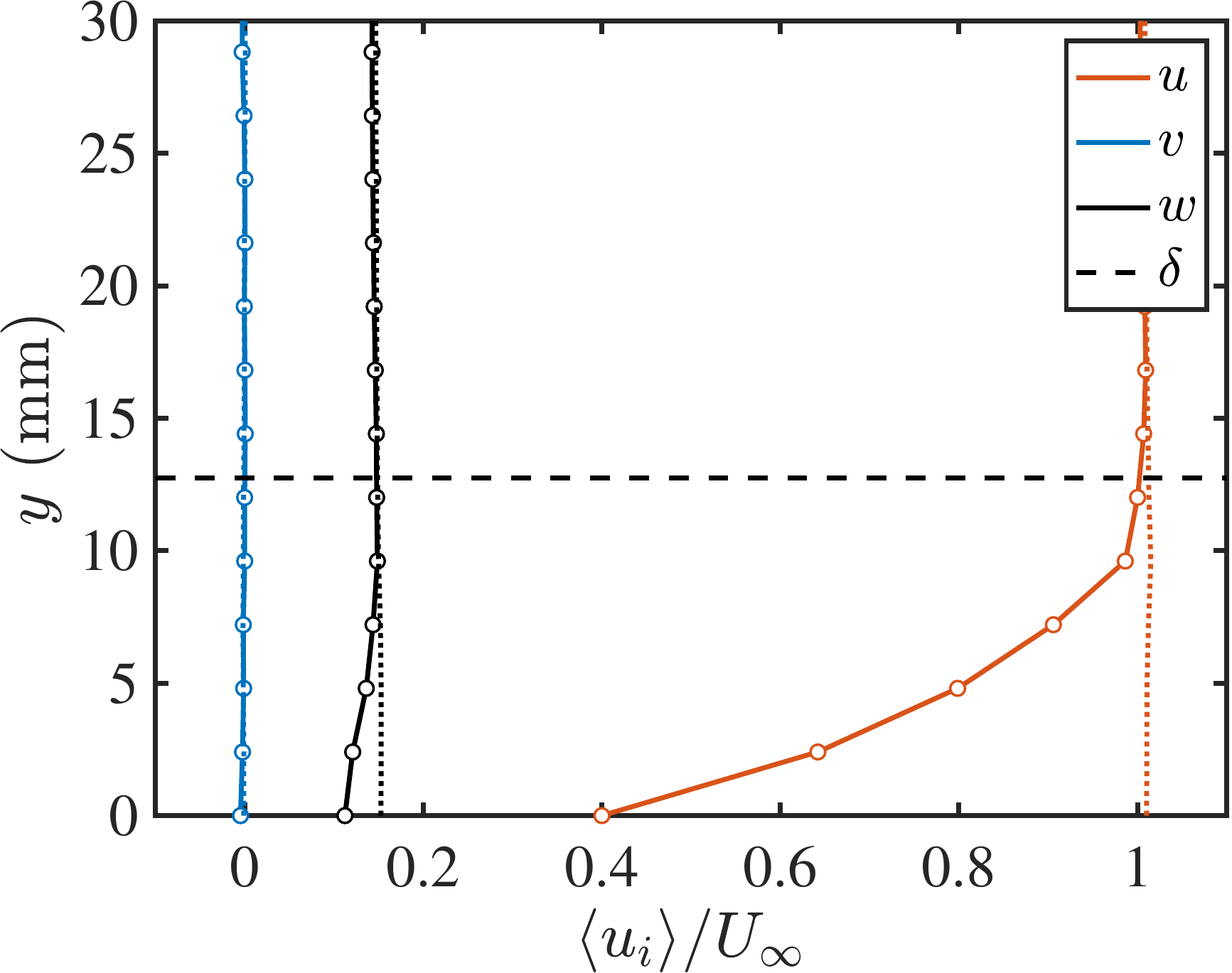}}
  \hspace{0.1cm}
  \subfloat[]{\includegraphics[width=0.47\textwidth]{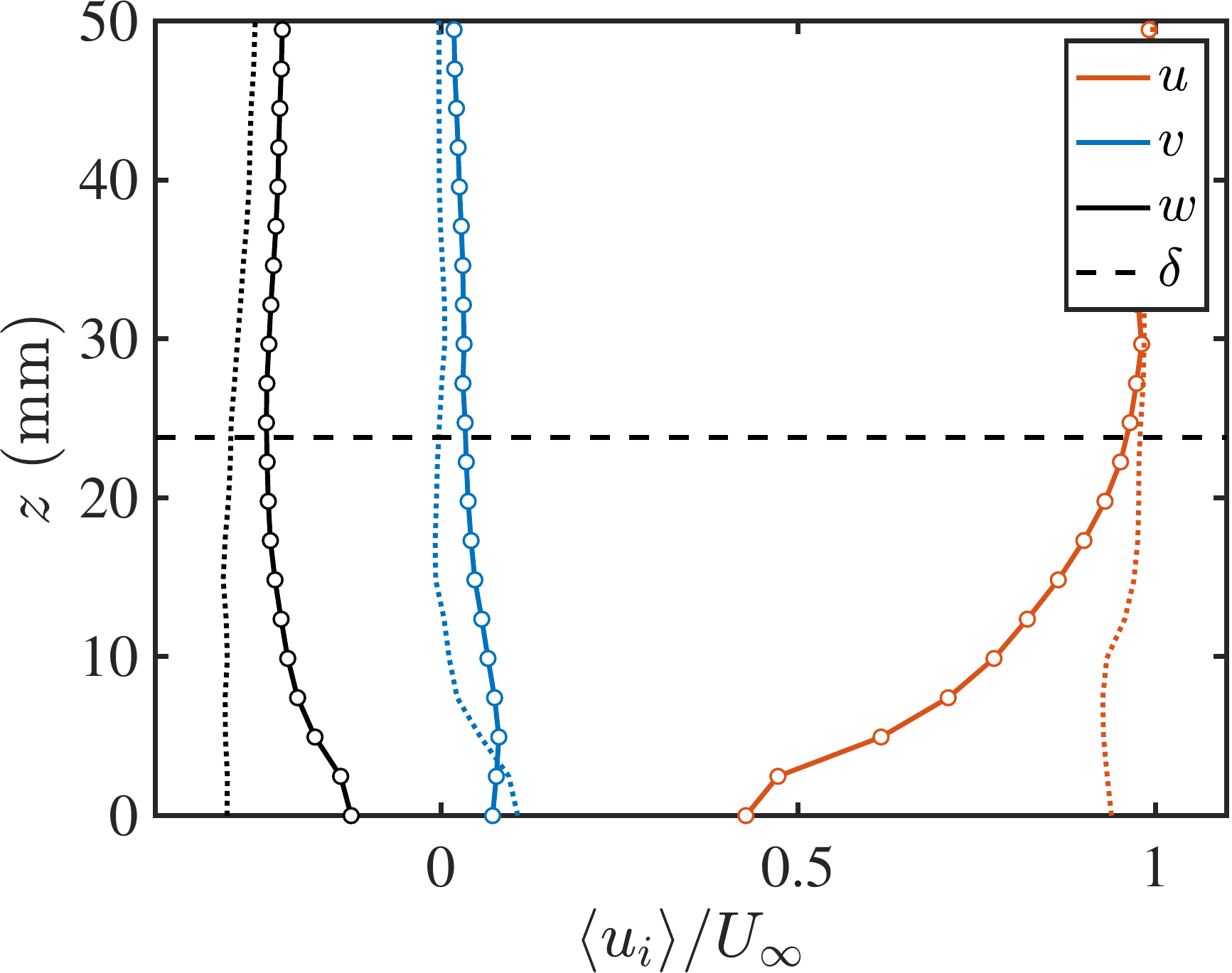}}
\end{center}
\begin{center}
  \includegraphics[width=0.8\textwidth]{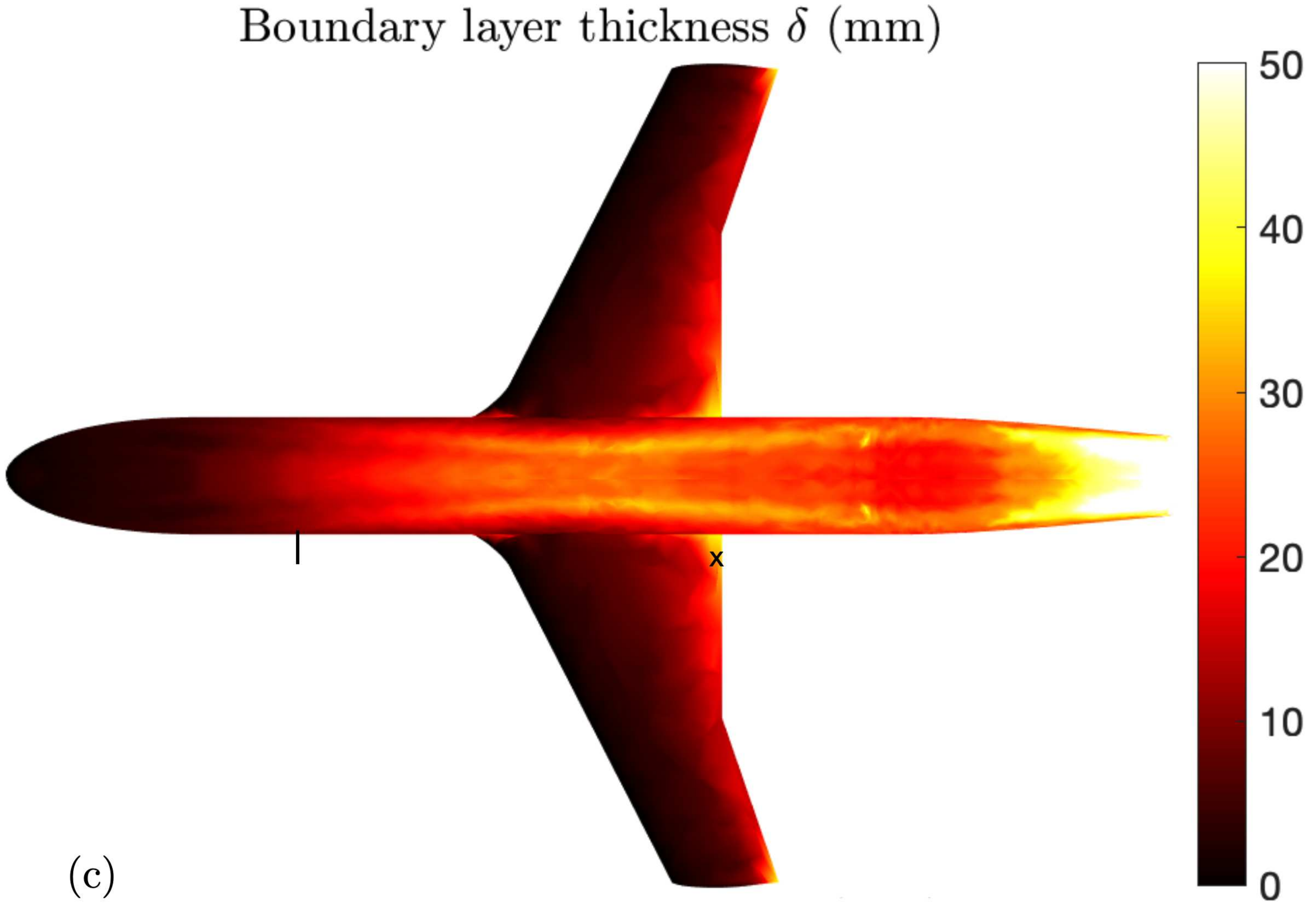}
\end{center}
\caption{The three mean velocity components for $\langle
  \boldsymbol{u}\rangle$ (lines with symbols) and $\langle
  \boldsymbol{u}_I\rangle$ (dotted lines), and boundary layer height
  (dashed). The locations of the mean profiles are indicated in panel
  (c) by the solid line for panel (a) and the cross for panel (b). (c)
  Boundary-layer thickness (in millimeters) for the NASA Juncture Flow
  at AoA = 5 degrees and $Re= 2.4 \times
  10^6$. \label{fig:tbl_thickness}}
\end{figure}

The gridding approach above requires an estimation of the
boundary-layer thickness at each location of the aircraft surface.
The method proposed here is based on measuring the deviation of the
viscous flow solution from the reference inviscid flow. This is
achieved by conducting two simulations: one WMLES, whose velocity is
denoted as $\boldsymbol{u}$, and one inviscid simulation (no SGS model
and free-slip at the wall), with velocity denoted by
$\boldsymbol{u}_I$. The grid generation of both simulations follows
strategy (i) with $\Delta =2$~mm. Boundary layers at the leading edge
with thickness below $2$~mm are estimated by extrapolating the
solution using a power law.  Two examples of mean velocity profiles
for $\boldsymbol{u}$ and $\boldsymbol{u}_I$ are shown in Figures
\ref{fig:tbl_thickness}(a) and (b). The three-dimensional surface
representing the boundary layer edge $S_\mathrm{tbl}$ is identified as
the loci of
  \begin{equation}\label{eq:Stbl}
    S_\mathrm{tbl} \equiv \left\{ (x,y,z)  : \frac{||\langle \boldsymbol{u}_I(x,y,z) \rangle
      - \langle \boldsymbol{u}(x,y,z) \rangle||}{|| \langle \boldsymbol{u}_I(x,y,z) \rangle||} = 0.01 \right\},
  \end{equation}
where $\langle \cdot \rangle$ denotes time-average. Finally, at each
point of the aircraft surface $(x_a,y_a,z_a)$, the boundary-layer
thickness $\delta$ is defined as the minimum spherical distance
between $\boldsymbol{x}_a = (x_a,y_a,z_a)$ and $\boldsymbol{x} =
(x,y,z) \in S_{tbl}$,
  \begin{equation}\label{eq:delta_def}
    \delta(\boldsymbol{x}_a) \equiv ||\boldsymbol{x}_a - \boldsymbol{x}||_\mathrm{min},
    \ \forall \boldsymbol{x} \in S_{tbl}.
  \end{equation}
The boundary-layer thickness for the flow conditions of the NASA
Juncture Flow is shown in Figure \ref{fig:tbl_thickness}(c) and ranges
from $0$~mm at the leading edge of the wing to $\sim 30$~mm at the
trailing edge of the wing. Thicker boundary layers about $50$~mm are
found in the downstream region of the fuselage. Equation
(\ref{eq:Stbl}) might be interpreted as the definition for a
turbulent/nonturbulent interface, although it also applies to laminar
regions. Other approaches for defining $S_{\mathrm{tbl}}$ were also
explored and combined with Eq. (\ref{eq:delta_def}), such as
isosurfaces of Q-criterion. Nonetheless, the present flow case is
dominated by attached boundary layers and Eq. (\ref{eq:Stbl}) yields
reasonable results for the purpose of generating a BL-conforming grid.
%

\subsection{Number of grid points}\label{subsec:number}

\begin{figure}
  \begin{center}
   \subfloat[]{\includegraphics[width=0.49\textwidth]{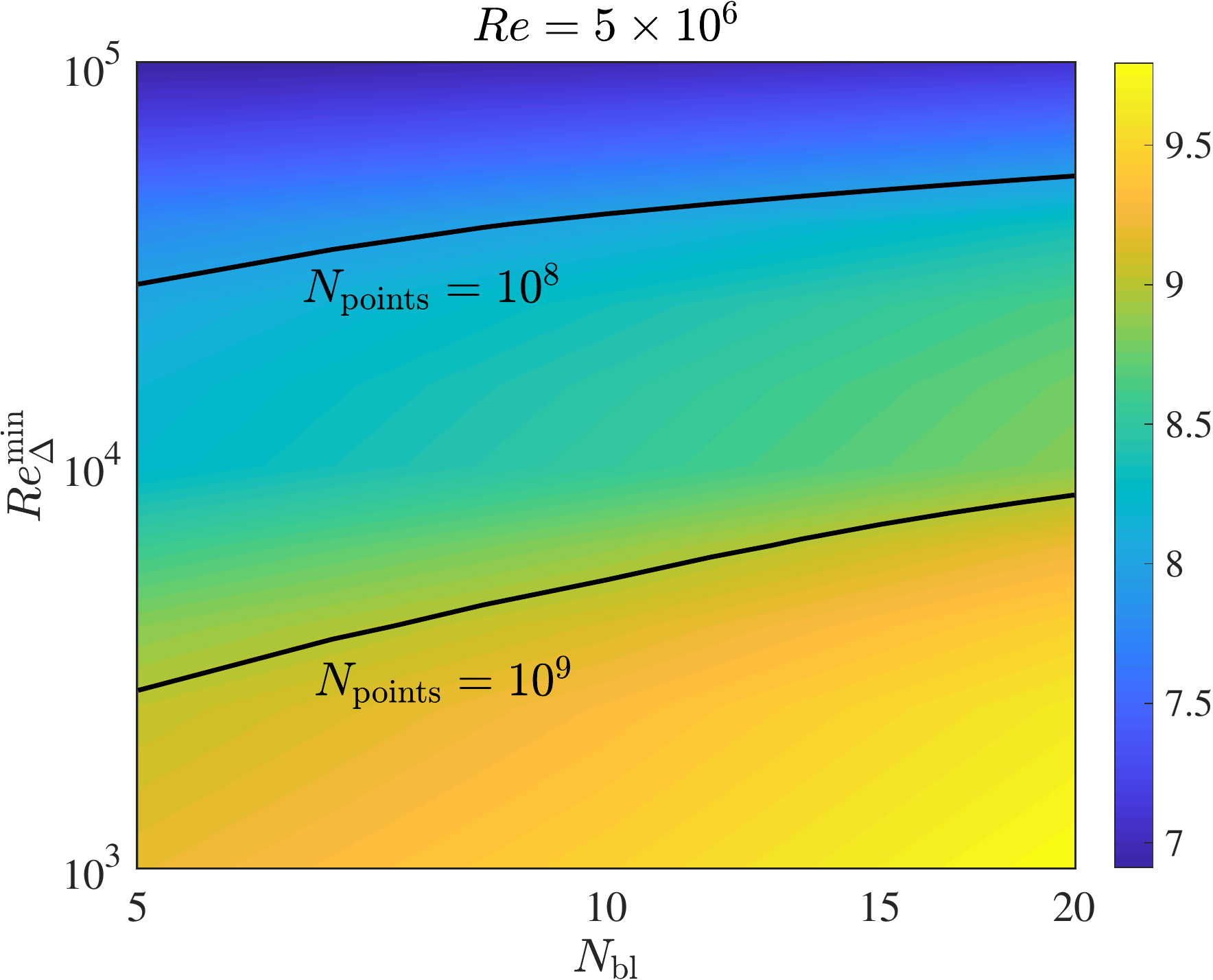}}
    \hspace{0.3cm}
   \subfloat[]{\includegraphics[width=0.44\textwidth]{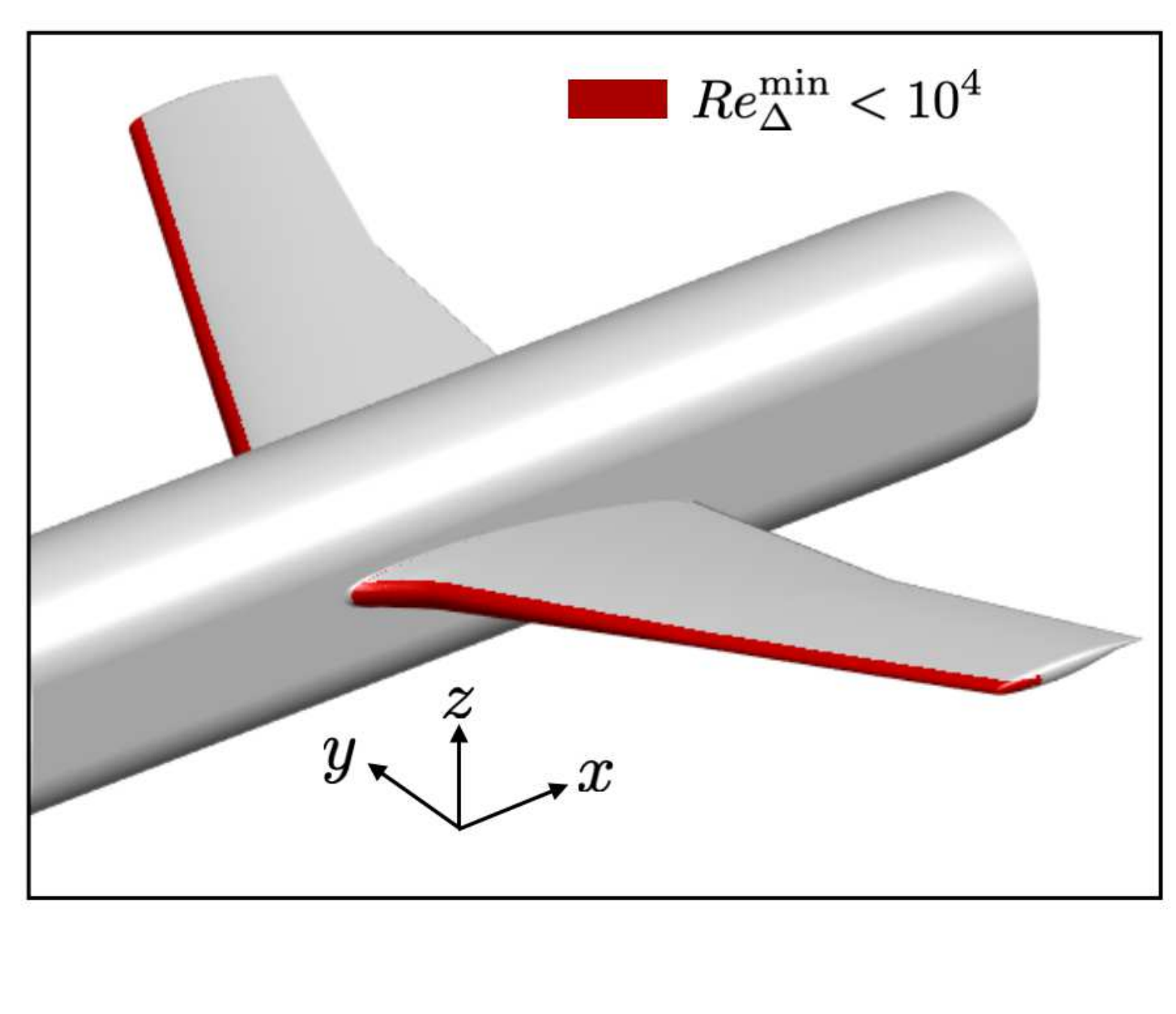}}
\end{center}
  \begin{center}
    \subfloat[]{\includegraphics[width=0.47\textwidth]{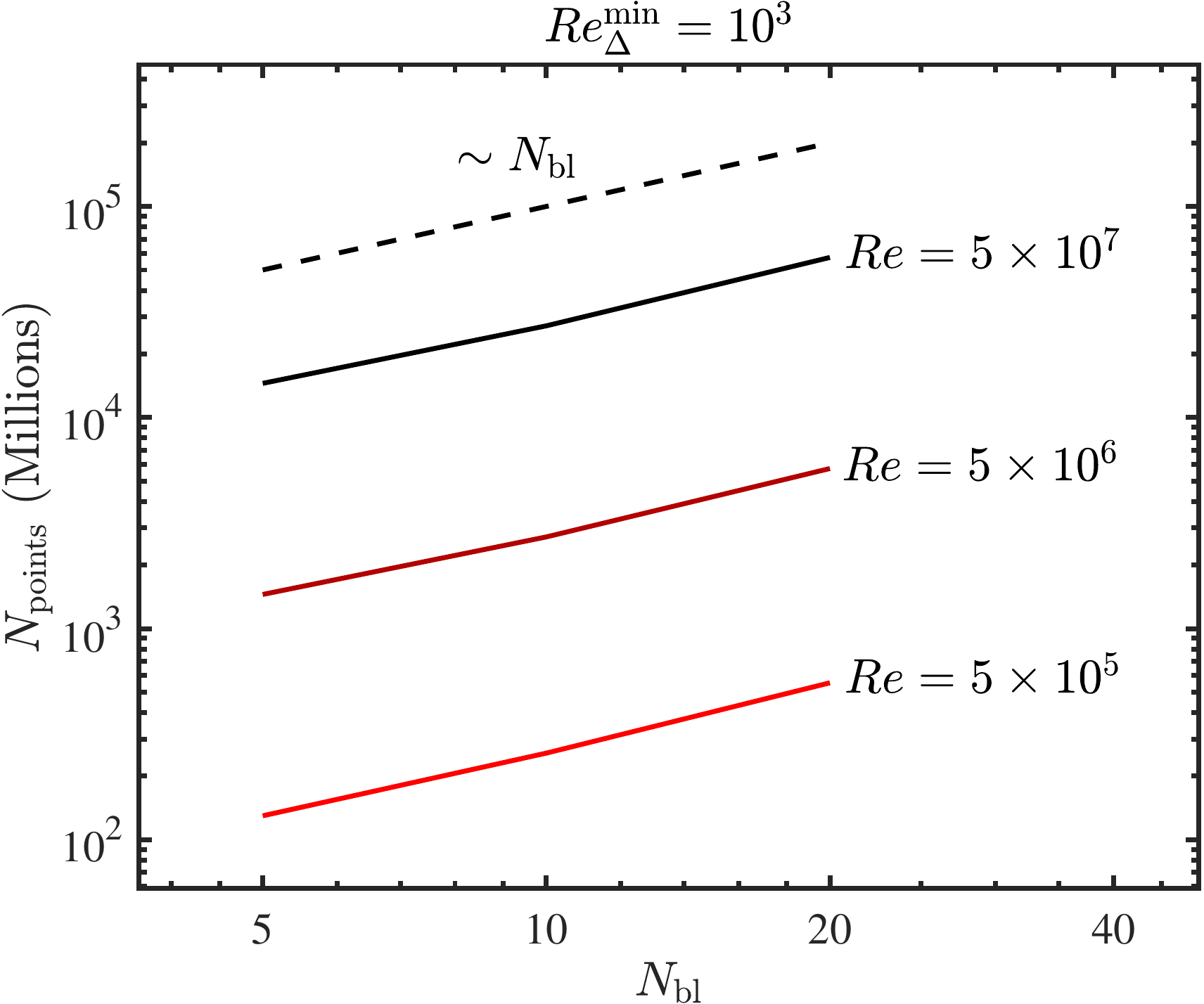}}
        \hspace{0.1cm}
  \subfloat[]{\includegraphics[width=0.47\textwidth]{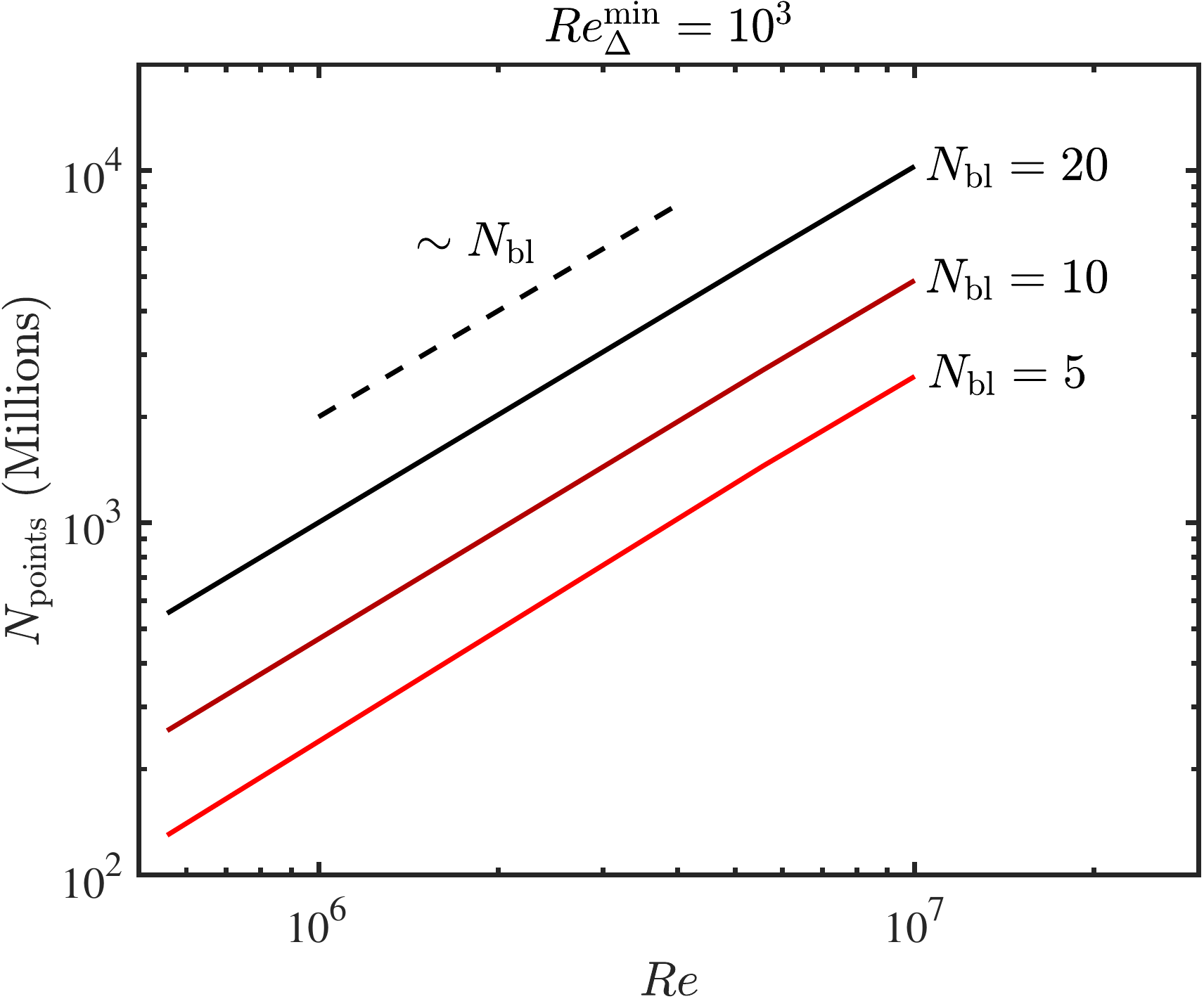}}
\end{center}
\caption{ (a) Logarithm of the number of points
  ($\log_{10}N_\mathrm{points}$) required for WMLES of the NASA
  Juncture Flow geometry as a function of the number of grid points
  per boundary-layer thickness ($N_{bl}$) and minimum grid Reynolds
  number ($Re_\Delta^\mathrm{min}$) for $Re = 5 \times 10^6$. (b)
  Subgrid boundary-layer region (in red) for $Re_\Delta^\mathrm{min} =
  10^4$ at $Re = 5 \times 10^6$. Panels (c) and (d) are the number of
  grid points as a function of (c) $N_{bl}$ and (d)
  $Re$.\label{fig:cost}}
\end{figure}

We estimate the number of grid points (or control volumes) to conduct
WMLES of the NASA Juncture Flow as a function of the number of points
per boundary-layer thickness ($N_{bl}$) and the minimum grid Reynolds
number ($Re_{\Delta}^\mathrm{min}$). We assume the gridding strategy
(ii) and utilize the Juncture Flow geometry. The boundary-layer
thickness was obtained following the procedure in Section
\ref{subsec:tbl}. The total number of points, $N_\mathrm{points}$, to
grid the boundary layer spanning the surface area of the aircraft
$S_a$ is
\begin{equation}\label{eq:points}
  N_\mathrm{points} = \int_{0}^{\delta} \int_{S_a} \frac{1}{\Delta(x_{||},y_{||})^3} \mathrm{d}x_{||} \mathrm{d}y_{||} \mathrm{d}n
  = \int_{S_a} \frac{N_{bl}}{\Delta(x_{\parallel},y_{\parallel})^2} \mathrm{d}x_{||} \mathrm{d}y_{||}, 
\end{equation}
where $x_{\parallel}$, $y_{\parallel}$ are the aircraft wall-parallel
directions, and $n$ is the wall-normal direction~\citep[see
  also][]{Chapman1979, Spalart1997, Choi2012}. Equation
(\ref{eq:points}) is integrated numerically and the results are shown
in Figure \ref{fig:cost}.  The cost map in Figure \ref{fig:cost}(a)
contains $\log_{10}(N_\mathrm{points})$ as a function of $N_{bl}$ and
$Re_{\Delta}^\mathrm{min}$. The accuracy of the solution is expected
to improve for increasing values of $N_{bl}$, i.e., higher energy
content resolved by the LES grid, and decrease with increasing
$Re_{\Delta}^\mathrm{min}$. The latter sets the minimum boundary-layer
thickness that can be resolved by the LES grid (i.e., the largest
thickness of the subgrid boundary layer).  Figure \ref{fig:cost}(b)
provides a visual illustration of the subgrid-boundary-layer region
for $Re_\Delta^\mathrm{min} < 10^4$, which is confined to a small
region (less than 10\% of the chord) at the leading edge of the wing.
%

The Reynolds number considered in Figure \ref{fig:cost}(a) is $Re=5
\times 10^6$, which is representative of wind tunnel experiments. A
Reynolds number typical of aircraft in flight conditions is
$Re\approx5 \times 10^7$, which would increase $N_\mathrm{points}$ by
a factor of ten due to the thinning of the boundary layers.  More
precisely, the increase in $N_\mathrm{points}$ is proportional to $Re$
and $N_{bl}$, and roughly inversely proportional to
$Re_\Delta^\mathrm{min}$.  The scaling properties of
Eq. (\ref{eq:points}) can be explained by assuming that the boundary
layer over the aircraft is fully turbulent and grows as $\delta \sim
(x-x_e)[(x-x_e)U_\infty/\nu]^{-m}$, where $x_e$ is the streamwise
distance to the closest leading edge and $m\approx-1/7$ for a zero-pressure
gradient flat-plate turbulent boundary layer (ZPGTBL). If we further
assume that $Re \gg Re_\Delta^\mathrm{min}$, the number of control
volumes can be shown to scale as
\begin{equation}\label{eq:growth}
N_\mathrm{points} \sim N_{bl} Re \left( Re_\Delta^\mathrm{min} \right)^{-5/6},
\end{equation}
which is confirmed in Figures \ref{fig:cost}(c) and (d) using the data
obtained using Eq. (\ref{eq:points}) for the NASA Juncture Flow.

The two isolines in Figure \ref{fig:cost}(a) bound the region
$N_\mathrm{points}=100$ million to $1000$ million grid points, which
are within the reach of current computing resources available to the
industry.  For example, for $N_{bl}\approx 10$ and
$Re_\Delta^\mathrm{min} \approx 10^4$, the required number of points
is $\sim 400$ million, which can be currently simulated in a few days
using thousands of cores. However, if the desired accuracy for the
quantities of interest is such that $N_{bl}\approx 20$ and
$Re_\Delta^\mathrm{min} \approx 10^3$, the number of grid points rises
up to $3000$ million, which renders WMLES unfeasible as a routine tool
for the industry. Hence, the key to the success of WMLES as a
designing tool resides in the accuracy of the solution achieved as a
function of $N_{bl}$ and $Re_\Delta^\mathrm{min}$. This calls for a
systematic error characterization of the quantities of interest, which
is the objective of the present preliminary work.

 \section{Error scaling of WMLES}
 \label{sec:results}

The solutions provided by WMLES are grid-dependent and multiple
computations are required in order to faithfully assess the quality of
results.  This raises the fundamental question of what is the
expected WMLES error as a function of the flow parameters and grid
resolution. Here, we follow the systematic error-scaling
characterization from \cite{Lozano2019a}. Taking the experimental
values ($q^\mathrm{exp}$) as ground truth, the error in the quantity
$\langle q \rangle$ can be expressed as
\begin{equation}\label{eq:error_general}
  \varepsilon_q \equiv \frac{||\langle q^\mathrm{exp}\rangle-\langle q \rangle||_n}{||\langle q^\mathrm{exp}\rangle||_n}
  = f\left( \frac{\Delta}{\delta}, Re, \mathrm{Ma},\mathrm{geometry},...\right),
\end{equation}
where $||\cdot||_n$ is the L$_2$-norm along the spatial coordinates of
$\langle q\rangle$, and the error function can depend on additional
non-dimensional parameters of the problem.  For a given geometry and
flow regime, the error function in Eq. (\ref{eq:error_general}) in
conjunction with the cost map in Figure \ref{fig:cost}(a) determines
whether WMLES is a viable approach in terms of accuracy and
computational resources available. For the NASA Juncture Flow, the
geometry, $Re$, and Ma are fixed parameters. If we further assume that
the error follows a power law, Eq. (\ref{eq:error_general}) can be
simplified as
\begin{equation}\label{eq:error_simply}
\varepsilon_q = c_q \left( \Delta/\delta \right)^{\alpha_q},
\end{equation}
where $c_q$ and $\alpha_q$ are constants that depend on the modeling
approach and flow region (i.e., laminar, fully turbulent,
separation,...). For turbulent channel flows, \cite{Lozano2019a}
showed that $\alpha_q \approx 1$ and $c_q$ is of the same order for
various SGS models.

We focus on the error scaling of pointwise time-averaged velocity
profiles and pressure coefficient, which are used as a proxy to
measure the quality of the WMLES solution.  From an engineering
viewpoint, the lift and drag coefficients are the most pressing
quantities of interest in aerodynamics applications. However, these
are integrated quantities which do not provide flow details and are
susceptible to error cancellation.  The granularity provided by
pointwise time-averaged quantities allows us to detect modeling
deficiencies and aids the development of new models. Unfortunately,
the pointwise friction coefficient is not available from the
experimental campaign, which hinders our ability to assess the
performance of the wall models more thoroughly.

\subsection{WMLES Cases and flow uncertainties}\label{Cases}

We perform WMLES of the NASA Juncture Flow with a leading-edge horn at
$Re=2.4 \times 10^6$ and AoA=$5^\circ$.  Seven cases are
considered. In the first six cases, we use grids generated using
strategy (i) with constant grid size in millimeters akin to the example
offered in Figure \ref{fig:tbl_grids}(a). In this case, the direct
impact of $Re_\Delta^{min}$ can be absorbed into $\Delta/\delta$ as is
done in Eq. (\ref{eq:error_simply}). The grid sizes considered are
$\Delta \approx 6.7, 4.3, 2.2, 1.1$ and $0.5$ millimeters, which are
labeled as C-D7, C-D4, C-D2, C-D1, and C-D0.5, respectively. Cases
C-D7, C-D4, and C-D2 are obtained by refining the grid across the
entire aircraft surface. For cases C-D1 and C-D0.5, the grid size is,
respectively, 1.1 and 0.5 millimeters only within a refinement box
along the fuselage and wing-body juncture defined by
$x\in[1770,2970]$~mm, $y\in[-300,-200]$~mm, and $z\in[-50,150]$~mm. An
additional case is considered to assess the impact on the accuracy of
BL-conforming Voronoi grids. The grid is generated using
strategy (ii) for $N_{bl} = \delta/\Delta = 5$ and
$Re_\Delta^\mathrm{min}=8\times 10^3$, as shown in Figure
\ref{fig:tbl_grids}(b). The case is denoted as C-N5-Rem8e3.

In the following, $\langle \cdot \rangle$ and $(\cdot)'$ denote
time-average and fluctuating component, respectively. For comparison
purposes, the profiles are interpolated to the grid locations of the
experiments. Statistical uncertainties in WMLES quantities are
estimated assuming uncorrelated and randomly distributed errors
following a normal distribution.  The uncertainty in $\langle q
\rangle$ is then estimated as $\Delta \langle q \rangle \equiv
\sigma/\sqrt{N_s}$, where $N_s$ is the number of samples for the
computing $\langle q \rangle$, and $\sigma$ is the standard deviation
of the samples.  The uncertainties for the mean velocity profiles and
pressure coefficient were found to be below 1\% and are not reported
in the plots.

\subsection{Mean velocity profiles and Reynolds stresses}
\label{subsec:vel_stress}

\begin{figure}
\begin{center}
\subfloat[]{\includegraphics[width=0.44\textwidth]{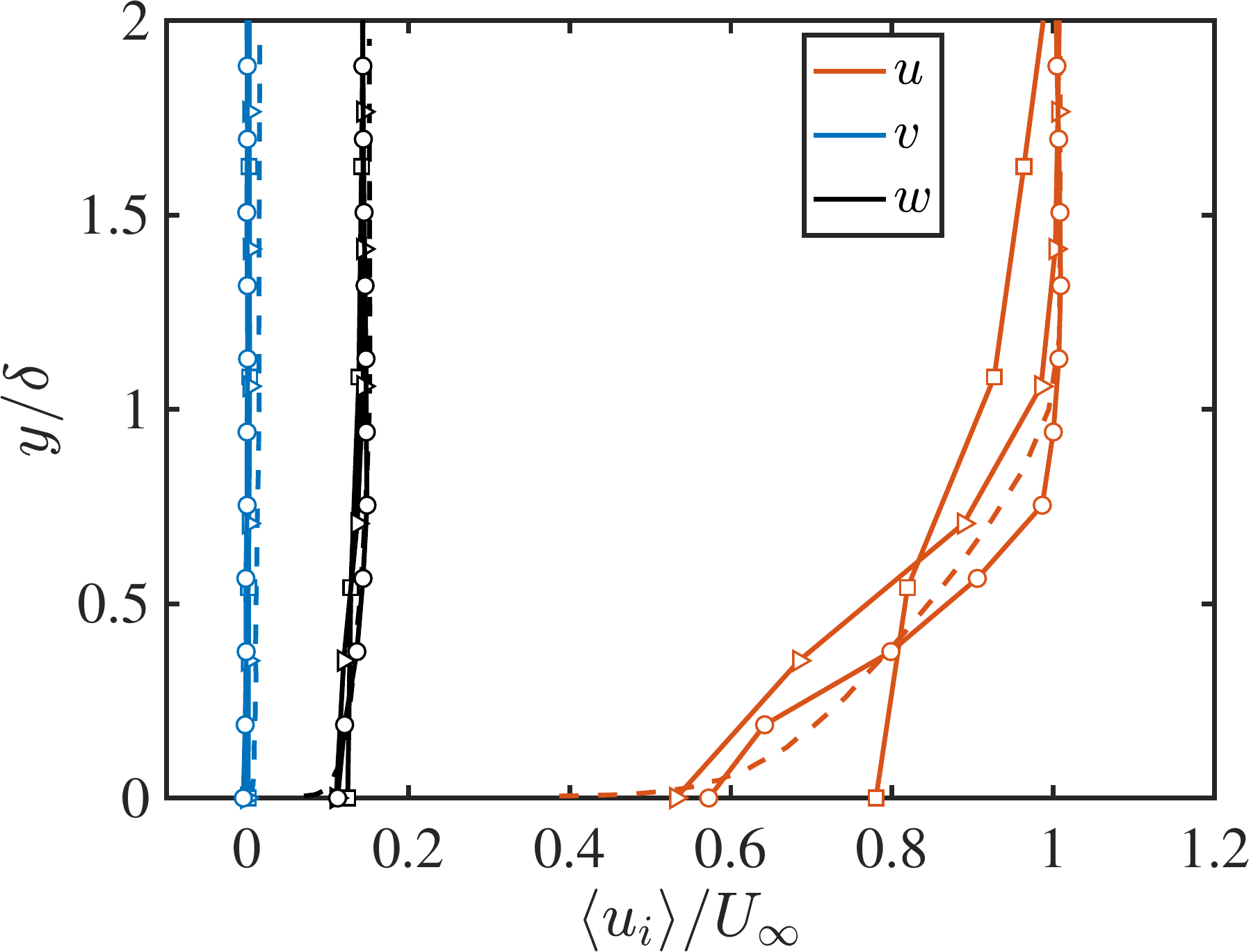}}
\hspace{0.5cm}
\subfloat[]{\includegraphics[width=0.47\textwidth]{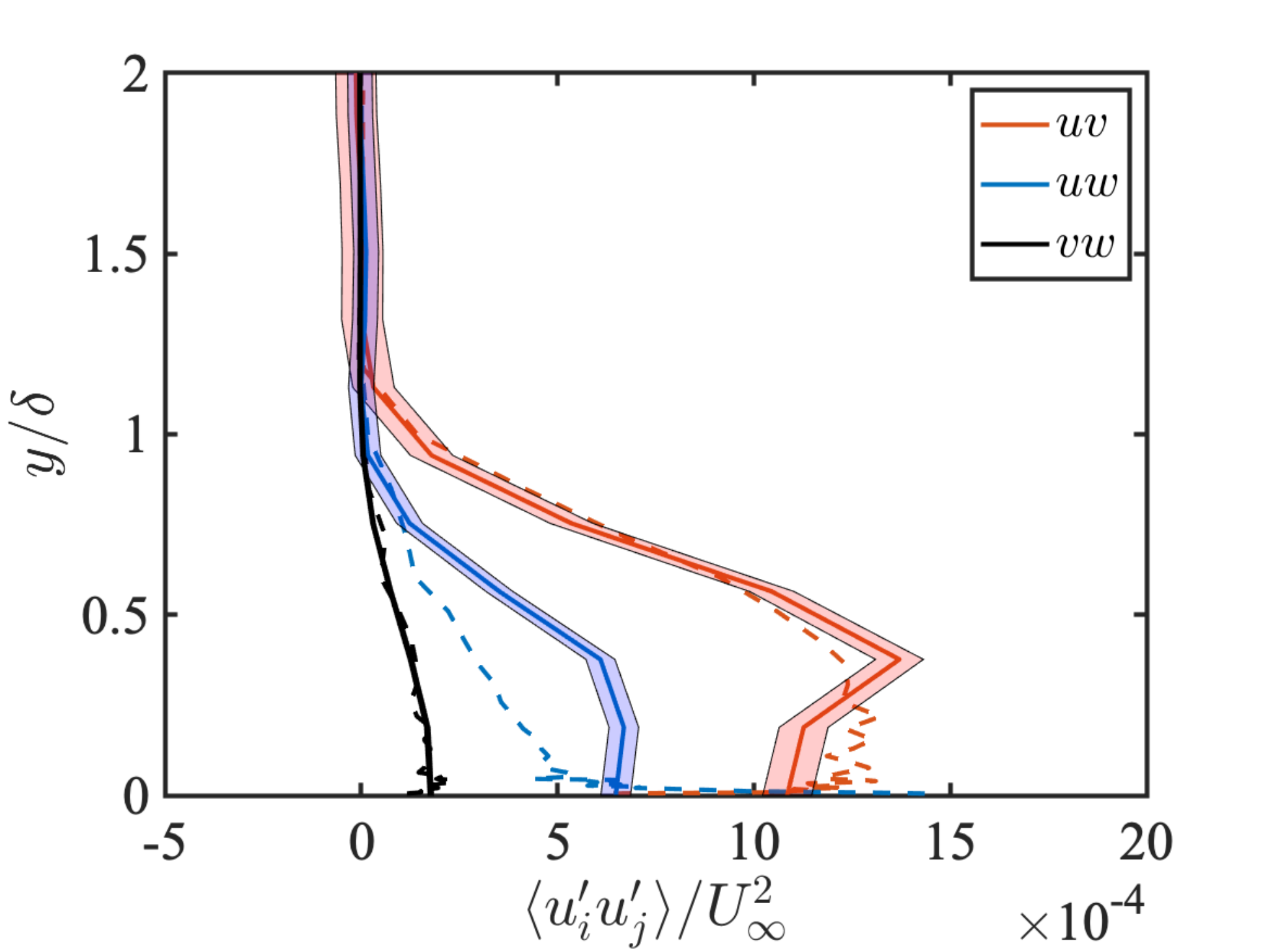}}
\end{center}
\caption{(a) Mean velocity profiles and (b) Reynolds stresses at
  location 1: upstream region of the fuselage $x=1168.4$~mm and
  $z=0$~mm (red line in Figure \ref{fig:Errors_all}(b)). Solid lines
  with symbols denote WMLES for cases C-D7 ($\square$), C-D4
  ($\triangleright$), and C-D2 ($\circ$). Colors denote different
  velocity components. Panel (b) only includes case C-D2 and the
  shaded area represents statistical uncertainty. Experiments are
  denoted by dashed lines. The distance $y$ is normalized by the local
  boundary-layer thickness $\delta$ at that
  location. \label{fig:fuselage}}
\end{figure}

\begin{figure}
\begin{center}
\subfloat[]{\includegraphics[width=0.44\textwidth]{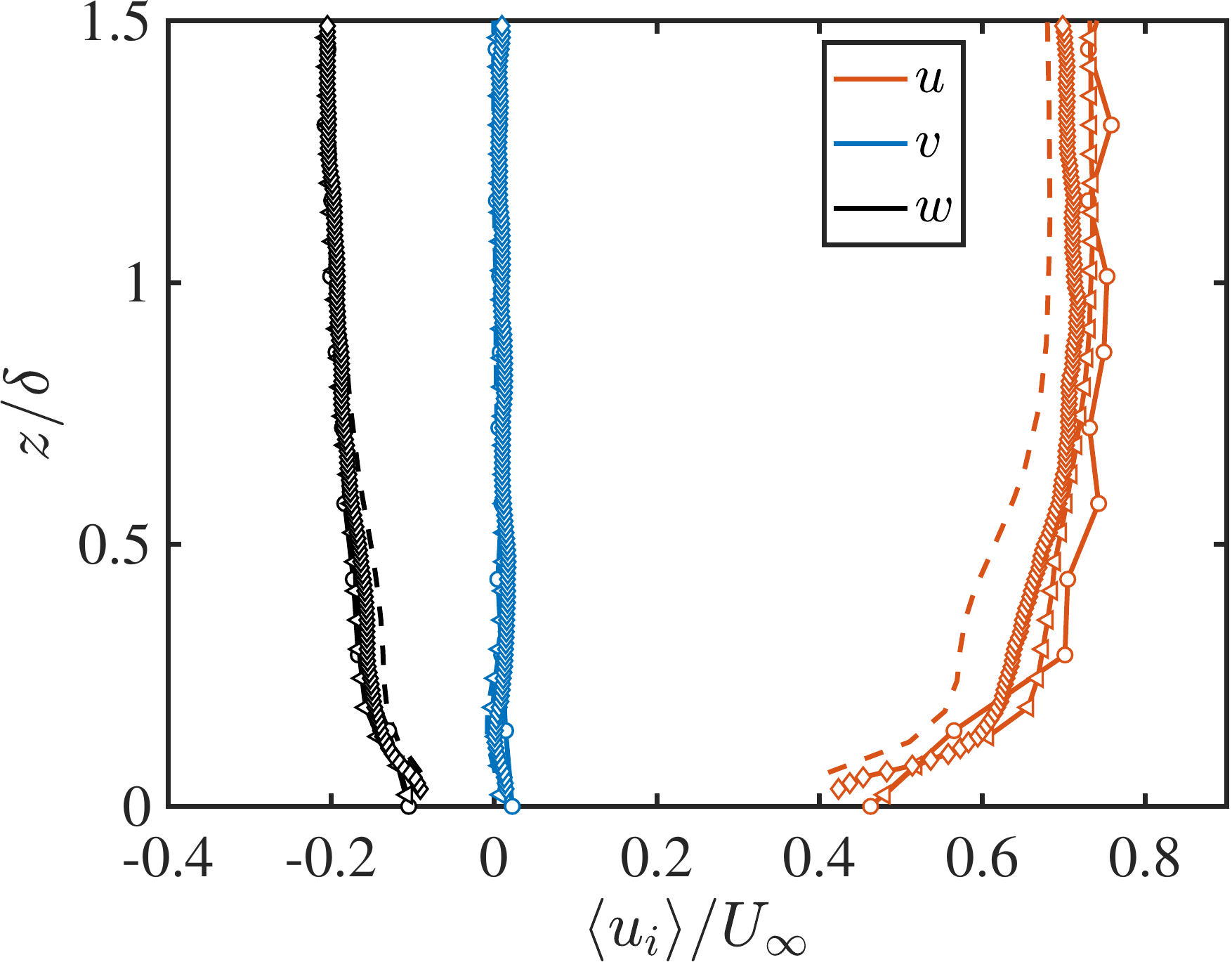}}
  \hspace{0.5cm}
\subfloat[]{\includegraphics[width=0.47\textwidth]{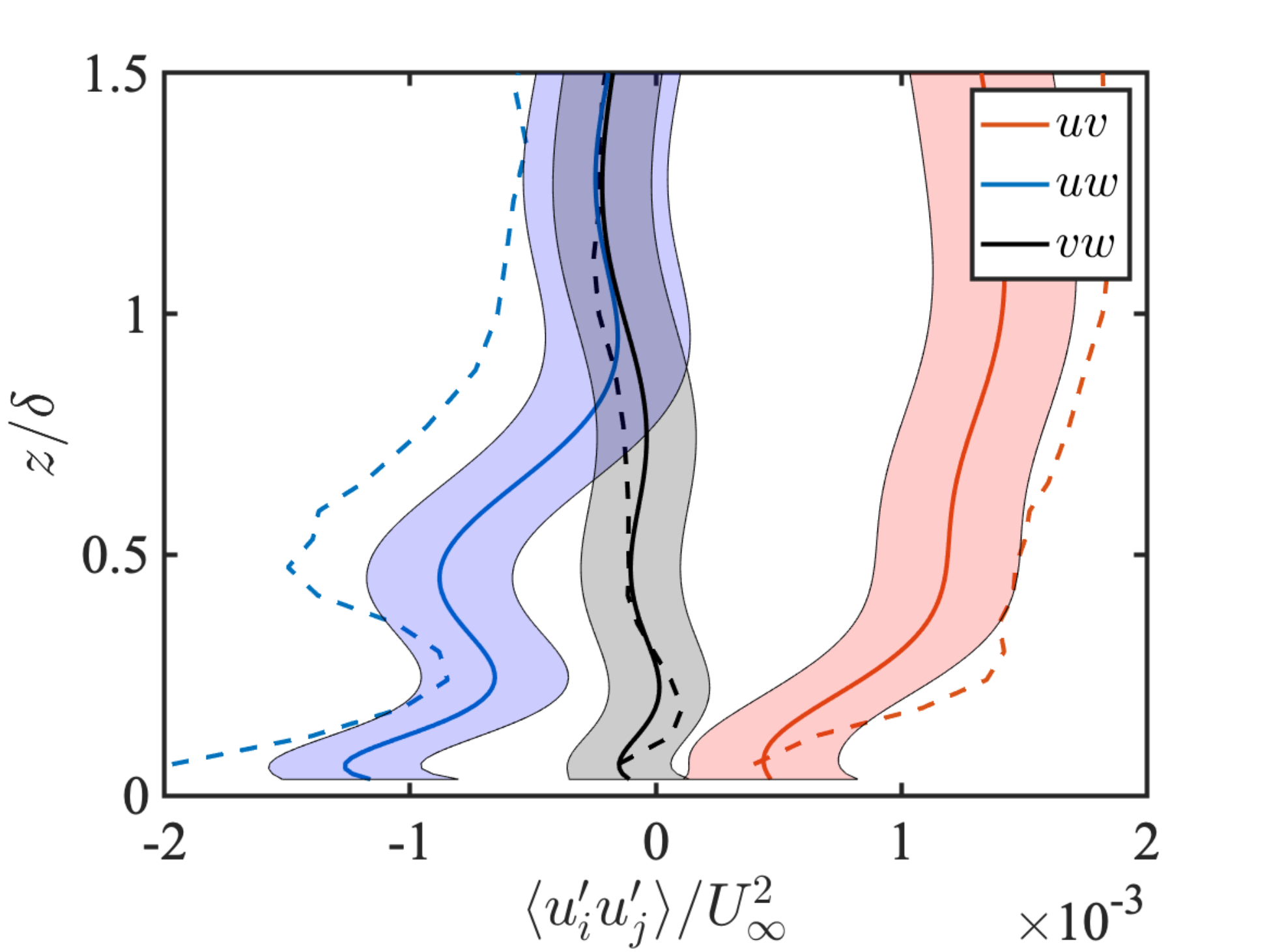}}
\end{center}
\caption{Same as Figure \ref{fig:fuselage} for location 2: wing-body
  juncture at $x=2747.6$~mm and $y=239.1$~mm (black line in Figure
  \ref{fig:Errors_all}(b)). In panel (a), lines with symbols are for
  cases C-D2 ($\circ$), C-D1 ($\triangleleft$), and C-D0.5
  ($\diamond$). In panel (b), the case shown in
  C-D0.5. \label{fig:juncture}}
\end{figure}

\begin{figure}
\begin{center}
\subfloat[]{\includegraphics[width=0.44\textwidth]{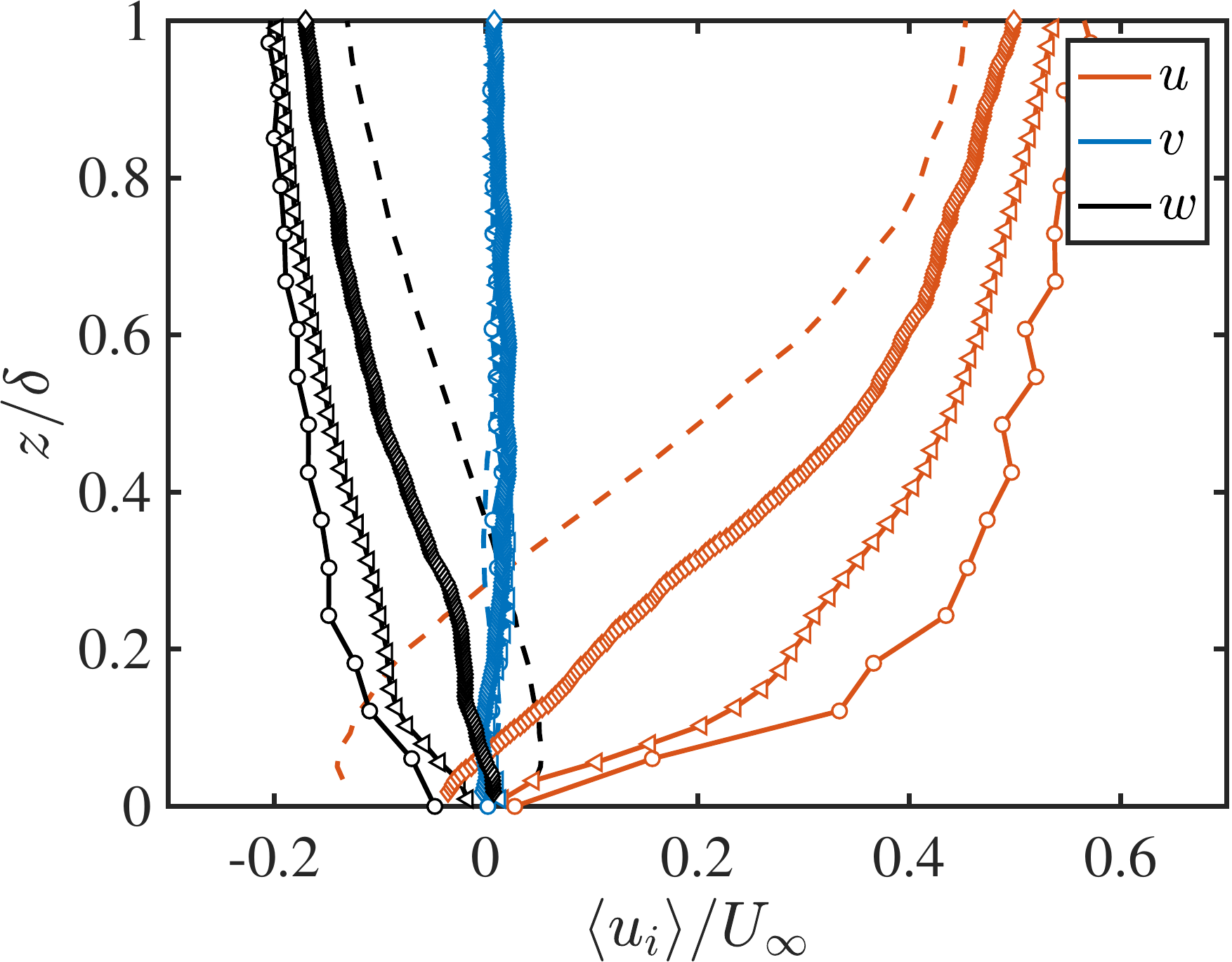}}
\hspace{0.5cm}
\subfloat[]{\includegraphics[width=0.47\textwidth]{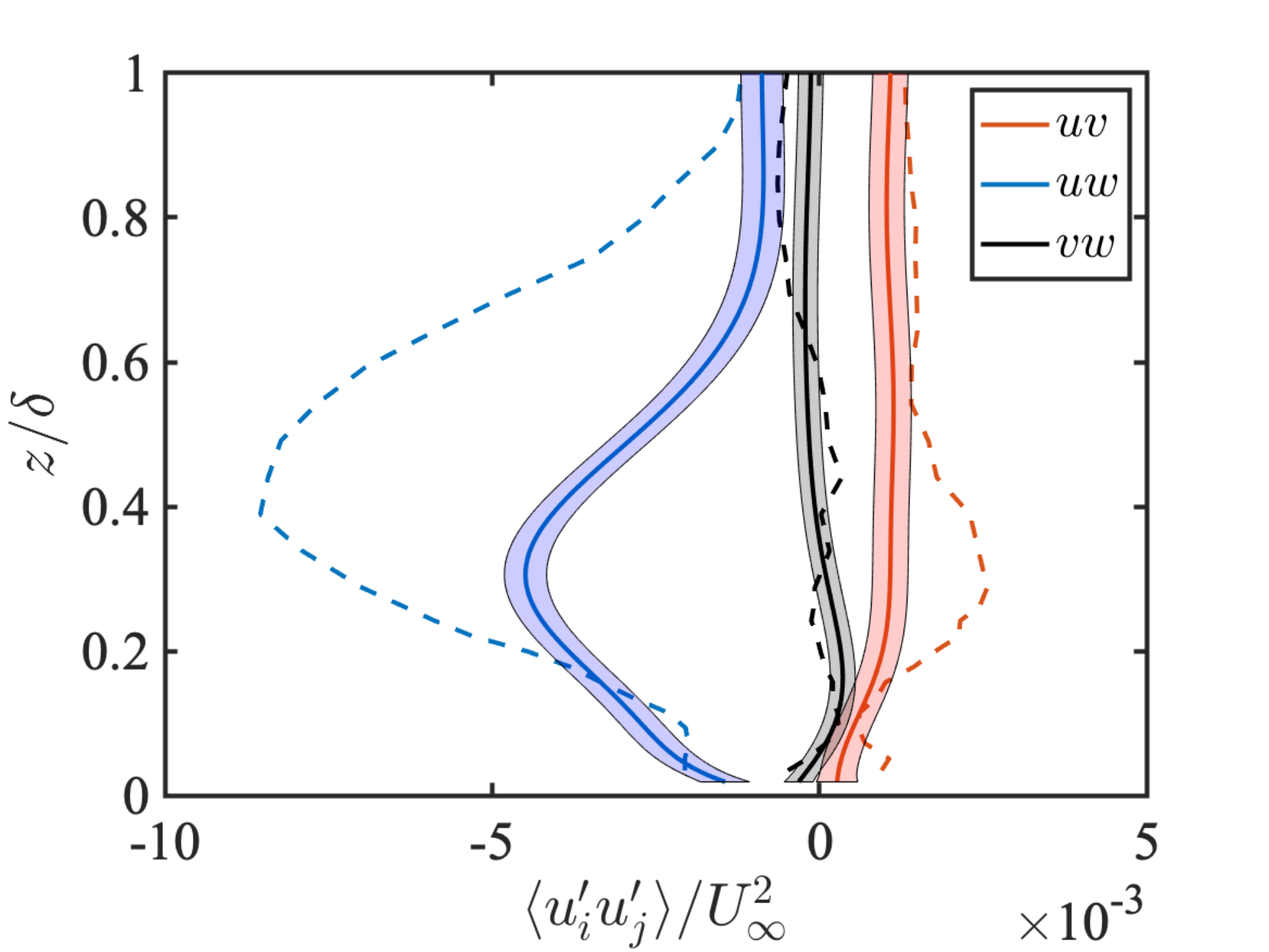}}
\end{center}
\caption{Same as Figure \ref{fig:fuselage} for location 3: wing-body
  juncture close to the trailing-edge at $x=2922.6$~mm and $y=239.1$~mm
  (blue line in Figure \ref{fig:Errors_all}(b)). In panel (a), lines
  with symbols are for cases C-D2 ($\circ$), C-D1 ($\triangleleft$),
  and C-D0.5 ($\diamond$). In panel (b), the case shown in
  C-D0.5.\label{fig:separation}}
\end{figure}

\begin{figure}
\begin{center}
\subfloat[]{\includegraphics[width=0.485\textwidth]{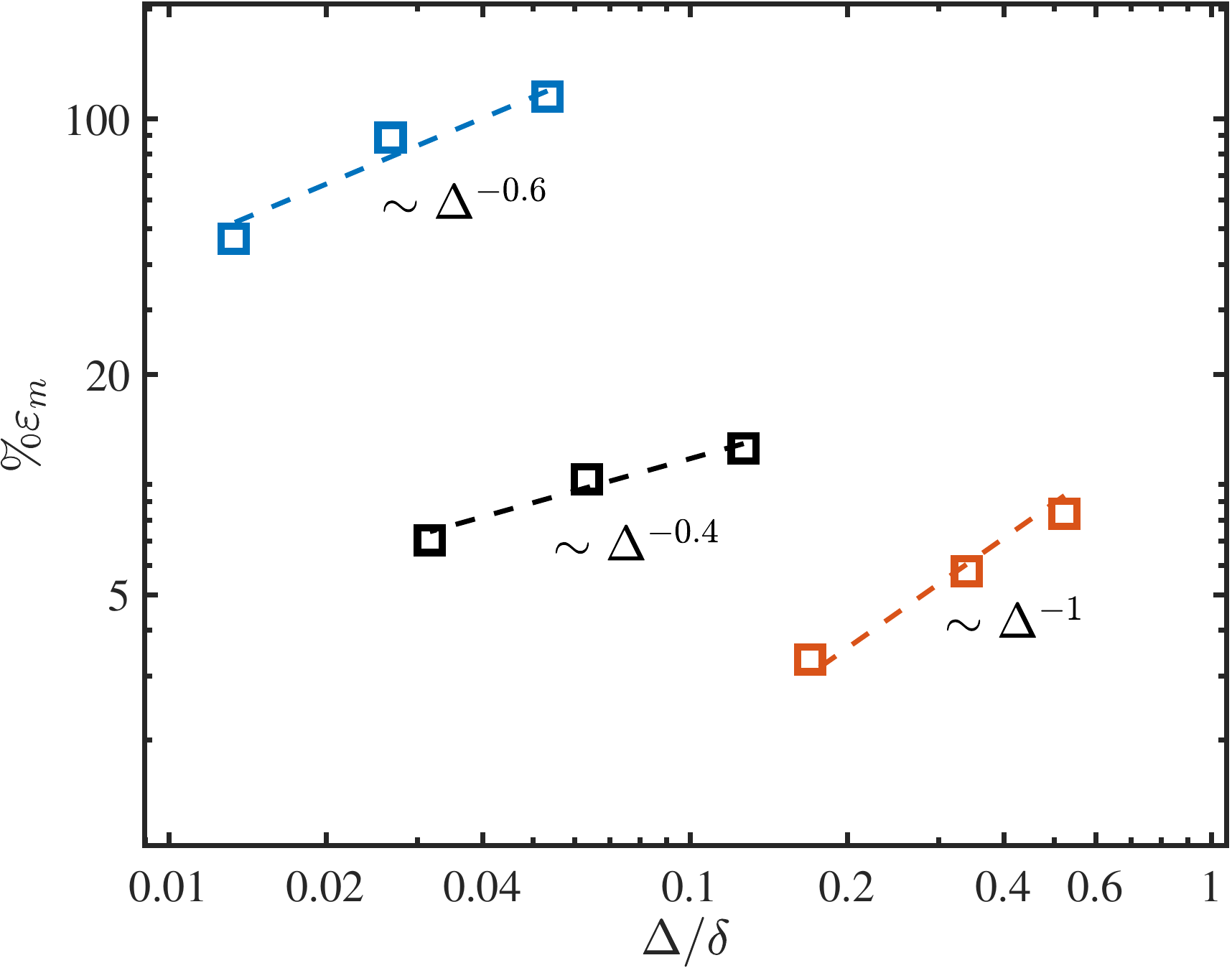}}
\subfloat[]{\includegraphics[width=0.46\textwidth]{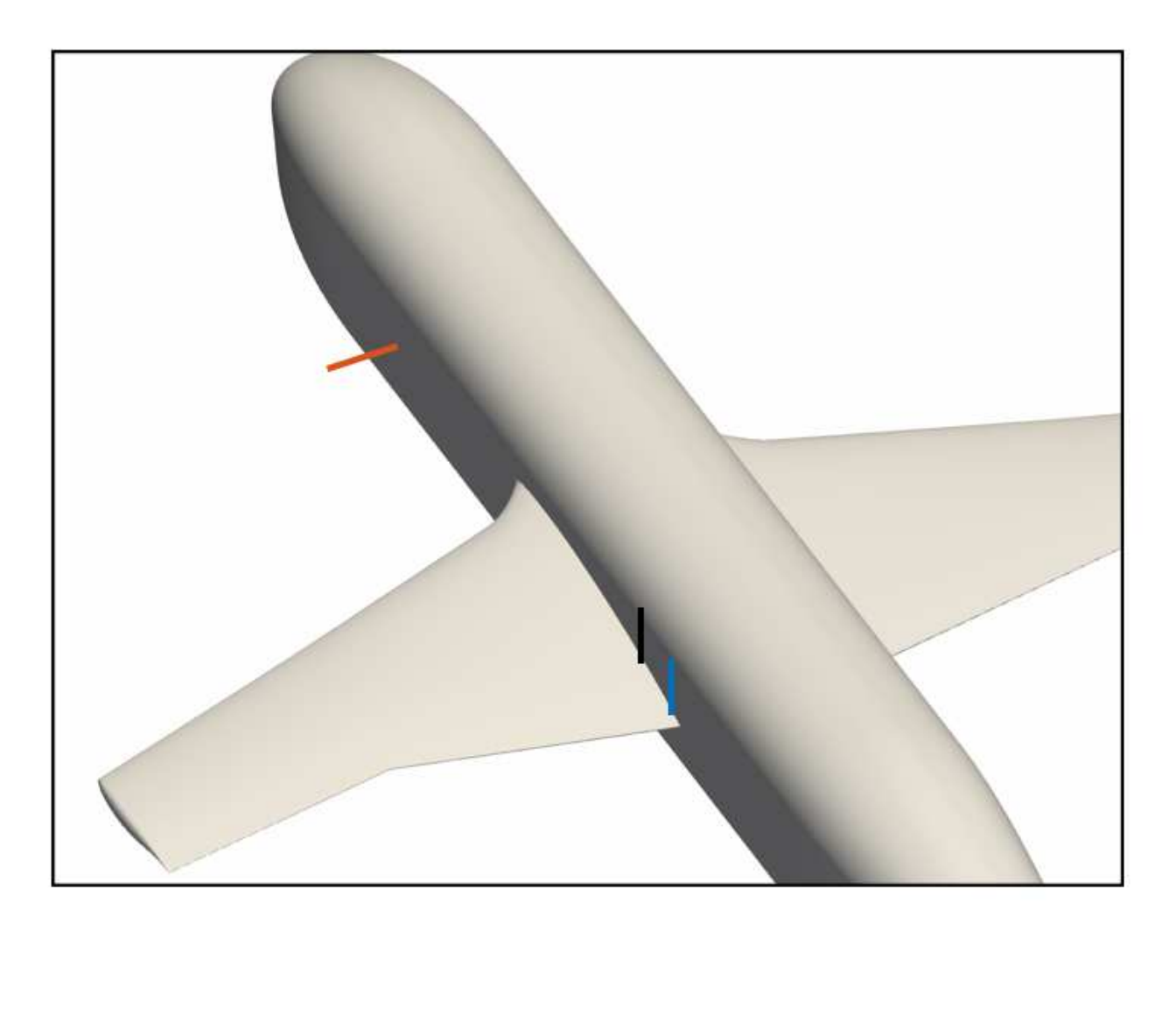}}
\end{center}
\caption{(a) Error in the mean velocity profile prediction by WMLES as
  a function of the grid resolution. The different colors denote the
  three locations indicated in panel (b) using the same color
  code.\label{fig:Errors_all}}
\end{figure}

We consider three locations on the aircraft: (1) the upstream region
of the fuselage, (2) the wing-body juncture, and (3) wing-body
juncture close to the trailing edge.  The mean velocity profiles are
shown in panel (a) of Figures \ref{fig:fuselage}, \ref{fig:juncture},
and \ref{fig:separation} for each location considered and the errors
are quantified in Figure \ref{fig:Errors_all}(a). We assume that
$\langle u_i \rangle \approx \langle u_i^{\mathrm{exp}} \rangle$ and
the mean velocity from WMLES is directly comparable with unfiltered
experimental data ($\langle u_i^{\mathrm{exp}} \rangle$). The
approximation is reasonable for quantities dominated by large-scale
contributions, as is the case for $\langle u_i \rangle$.  Figure
\ref{fig:fuselage}(a) shows that $\langle u_i \rangle$ from WMLES
converges to the experimental results with grid refinement.  The
turbulent boundary layer over the fuselage is about $10$ to $20$~mm
thick, which yields roughly 3--6 points per boundary-layer thickness
at the grid resolutions considered.  Assuming that the flow at a given
station can be approximated by a local canonical ZPGTBL, the expected
error in the mean velocities can be estimated as $\varepsilon_m
\approx 0.16 \Delta/\delta$~\citep{Lozano2019a}. For the current
$\Delta$, this yields errors of 2\%--8\%, consistent with the results
in Figure \ref{fig:Errors_all}(a) (red symbols).  The situation
differs for the wing-body juncture and wing trailing edge. Despite the
finer grid sizes, errors in the mean flow prediction are about 15\% in
the juncture (black symbols in Figure \ref{fig:Errors_all}a) and even
100\% in the trailing edge (blue symbols in Figure
\ref{fig:Errors_all}a).  The larger errors may be attributed to the
presence of a three-dimensional boundary layers and flow separation in
the vicinity of the wing-body juncture and trailing edge, which makes
the error scaling predicated upon the assumption of local similarity
to a ZPGTBL inappropriate.
%
%
%

The resolved portion of the tangential Reynolds stresses is shown in
Figure panels (b) of Figures \ref{fig:fuselage}, \ref{fig:juncture},
and \ref{fig:separation}.  The trends followed by $\langle u_i'
u_j'\rangle$ are correctly captured at the different stations
investigated, although their magnitudes tend to be systematically
underpredicted, especially for the juncture region and trailing edge.
Estimations from \cite{Lozano2019a} suggest that the error for the
turbulence intensities should scale as $\sim (\Delta/\delta)^{2/3}$,
which for the present grid resolution implies $\sim10$--$20\%$
error. The result is consistent with the typical $\Delta$ for WMLES,
which supports a limited fraction of the turbulent fluctuations.
Assuming $\langle u_i^\mathrm{exp} \rangle \approx \langle
u_i\rangle$, then $\langle {u_i'}^\mathrm{exp}
{u_j'}^\mathrm{exp}\rangle \approx \langle u_i' u_j'\rangle + \langle
\tau_{ij}^\mathrm{SGS}\rangle$, where $\tau_{ij}^\mathrm{SGS}$ is the
subgrid-scale tensor.  Thus, for the current grid sizes we can expect
$|\langle u_i' u_j'\rangle| < |\langle {u_i'}^\mathrm{exp}
{u_j'}^\mathrm{exp}\rangle|$, i.e., severe underprediction of the
tangential Reynolds stress by WMLES.

Figure \ref{fig:Errors_all}(a) is the cornerstone of the present
study. It shows the relative errors in the prediction of the mean
velocity profiles for the three regions considered: upstream fuselage,
wing-body juncture, and wing-body juncture at the trailing edge. The
three regions are marked with lines in Figure \ref{fig:Errors_all}(b)
using the same colors as in Figure \ref{fig:Errors_all}(a).  The
turbulent flow in the fuselage resembles a ZPGTBL. As such, the wall-
and SGS models, which have been devised for and validated in
flat-plate turbulence, perform accordingly.  On the contrary, there is
a clear decline of current models in the wing-body juncture and
trailing edge region, which are dominated by secondary motions in the
corner and flow separation. Not only is the magnitude of the errors
larger in the latter locations, but the error rate of convergence is
also slower ($\varepsilon_m\sim(\Delta/\delta)^{0.5}$) when compared
with the equilibrium conditions encountered in ZPGTBL
($\varepsilon_m\sim(\Delta/\delta)$). The situation could be even more
unfavorable, as \cite{Lozano2019a} has shown that for refining grid
resolutions the convergence of WMLES towards the DNS solution may
follow a non-monotonic behavior due to the interplay between numerical
and SGS model errors.
%

\subsection{Pressure coefficient}\label{subsec:pressure}

\begin{figure}
\begin{center}
\includegraphics[width=1.\textwidth]{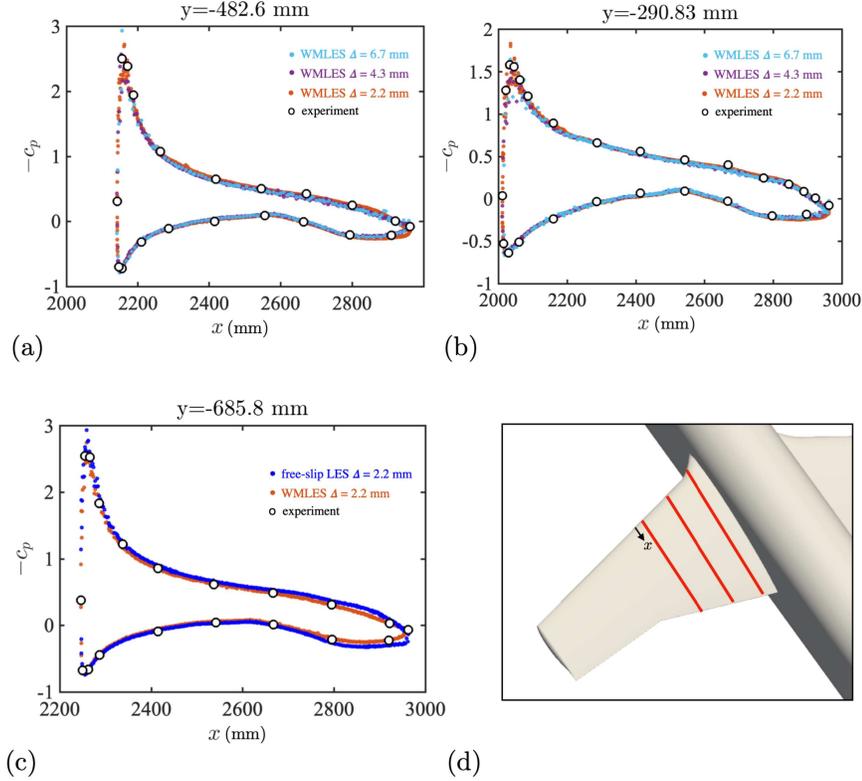}
\end{center}
\caption{ The surface pressure coefficient $C_p$ along the
  wing. Panels (a) and (b) show $C_p$ for cases C-D7, C-D4 and
  C-D2. Panel (c) shows $C_p$ for case C-D2 and a case identical to
  C-D2 but imposing free-slip boundary condition at the walls. (d)
  Locations over the wing selected to represent $C_p$ in panels (a),
  (b) and (c). \label{fig:Cp}}
\end{figure}

The surface pressure coefficient over the wing, $C_p$, is shown in
Figure \ref{fig:Cp} along the chord of the wing. The predictions are
compared with experimental data at three different $y$-locations. The
locations selected are denoted by red lines in Figure
\ref{fig:Cp}(d). Overall, WMLES agrees with the experimental data to
within 1--5\% error. The predictions are still to within 5\% accuracy
even at the coarsest grid resolutions considered, which barely
resolved the boundary layer. The main discrepancies with experiments
are located at the leading edge of the wing.

The accurate prediction of $C_p$ is a common observation in CFD of
external aerodynamics. The result can be attributed to the outer-layer
nature of the mean pressure, which becomes less sensitive to the
details of near-wall turbulence.  Under the thin boundary-layer
assumption, the integrated mean momentum equation in the spanwise
direction shows that $\langle p \rangle + \rho \langle v^2 \rangle
\approx p_e(x) \Rightarrow p_\mathrm{wall} = p_e(x)$, where $p_e$ is
the far-field pressure.  Hence, the pressure at the surface is mostly
imposed by the inviscid imprint of the outer flow.  This is demonstrated
by performing an additional calculation similar to C-D2 but imposing
the free-slip boundary condition at the wall such that boundary layers
are unable to develop (Figure \ref{fig:Cp}(c)). The conditions for
$p_\mathrm{wall} = p_e(x)$ would not hold, for example, when the wall
radius of curvature is comparable to the boundary-layer
thickness. This is the case in the vicinity of the wing leading-edge,
which is the region where the accuracy of $C_p$ provided for WMLES is
the poorest and most sensitive to $\Delta$. The tripping methodology
used in WMLES differs from the experimental setup, which may also
contribute to the discrepancies observed.

The outer-flow character of $C_p$ is encouraging for the prediction of
the pressure-induced components of the lift and drag coefficients.  It
also suggests that $C_p$ might not be a challenging quantity to
predict in the presence of wall-attached boundary layers.  Thus, the
community should focus the efforts on other important quantities of
interest such as the numerical prediction of the skin-friction
coefficient and its challenging experimental measurement.
%

\subsection{Separation bubble}\label{subsec:bubble}

\begin{figure}
\centering
\includegraphics[width=0.8\textwidth]{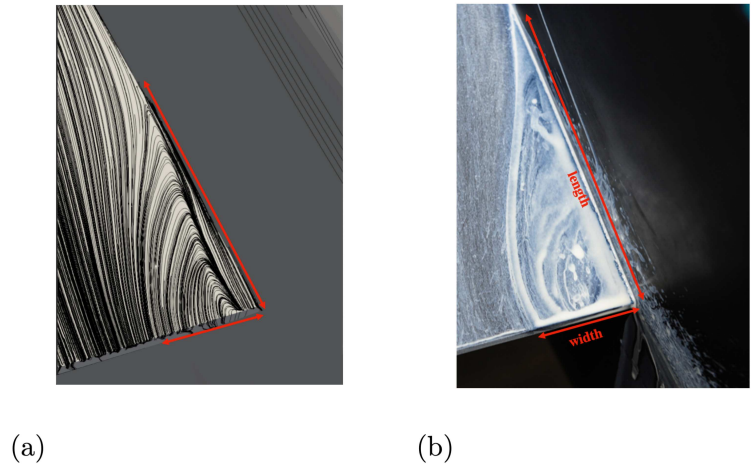}
\caption{Streamlines of the average tangential wall-stress. The
  results are for case C-D0.5. \label{fig:bubble_tau}}
\end{figure}

For completeness, we also report the results on the size of the
separation bubble. The mean wall-stress streamlines for case C-D0.5
are shown in Figure \ref{fig:bubble_tau}. The figure also contains a
depiction of the average length and width of the separation bubble,
which are about $100$ mm and $60$ mm, respectively, for case C-D0.5.
Direct comparison of these dimensions with oil-film experimental
results show that the current WMLES prediction is about 15\% lower
than the experimental measurements ($120 \times 80$ mm), consistent
with previous WMLES investigations~\citep{Lozano_AIAA_2020, Iyer2020,
  Ghate2020}. Nonetheless, note that the sizes of the separation zone
from WMLES are obtained from the tangential wall-stress streamlines
after the wall stress is averaged in time, whereas the experimental
sizes are obtained from the pattern resulting from the oil-film time
evolution. Albeit both methodologies provide an average description of
the size of the separation zone, they do not allow for a one to one
comparison and we should not interpret the present differences as a
faithful quantification of the errors. Hence, a methodology allowing
for quantitative comparisons of separated-flow patterns between CFD
and experimental data remains an open challenge.
%

\subsection{Improvements with boundary-layer-conforming grids}
\label{subsec:targeted}

\begin{figure}
\begin{center}
  \subfloat[]{\includegraphics[width=0.32\textwidth]{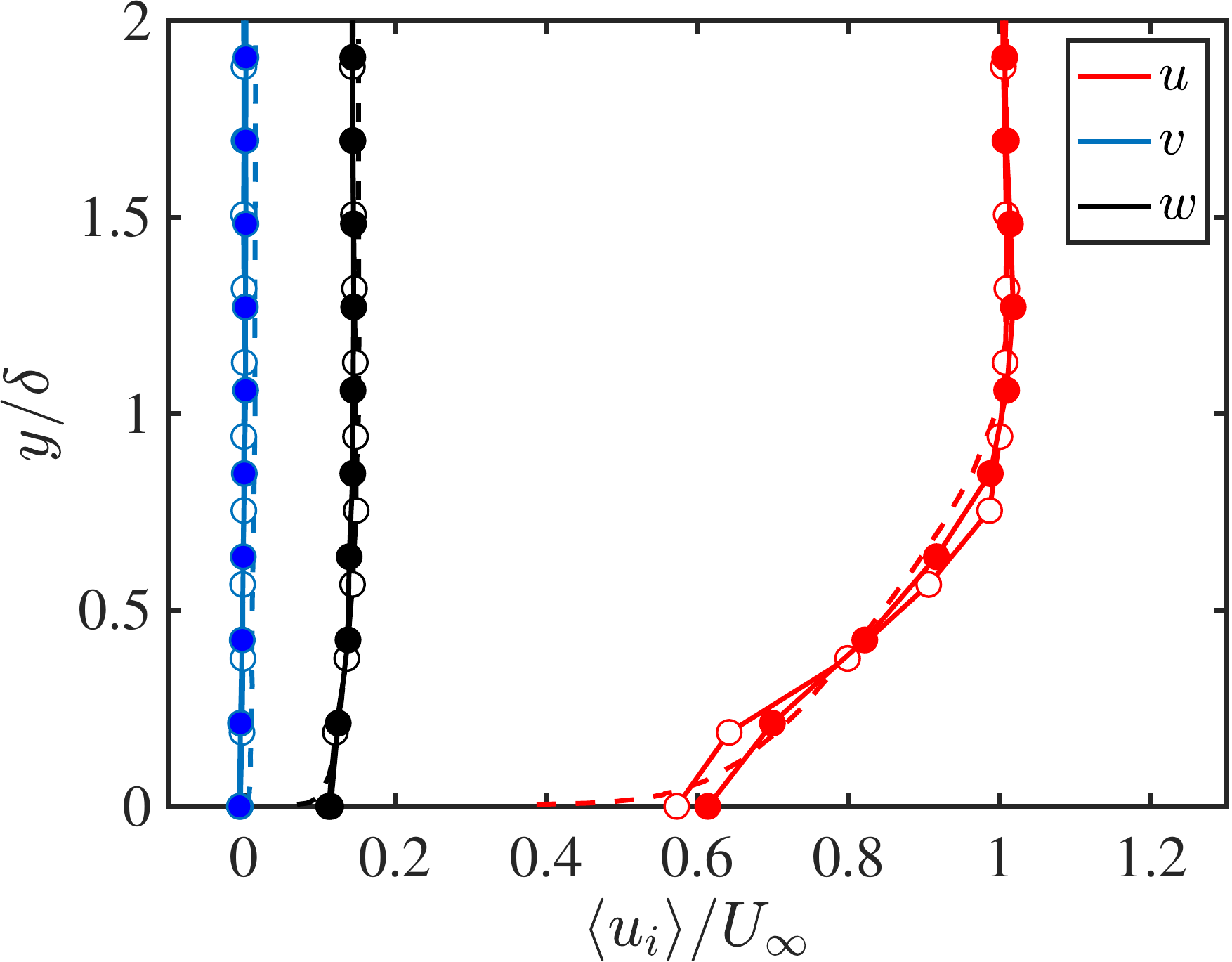}}
  \hspace{0.05cm}
  \subfloat[]{\includegraphics[width=0.32\textwidth]{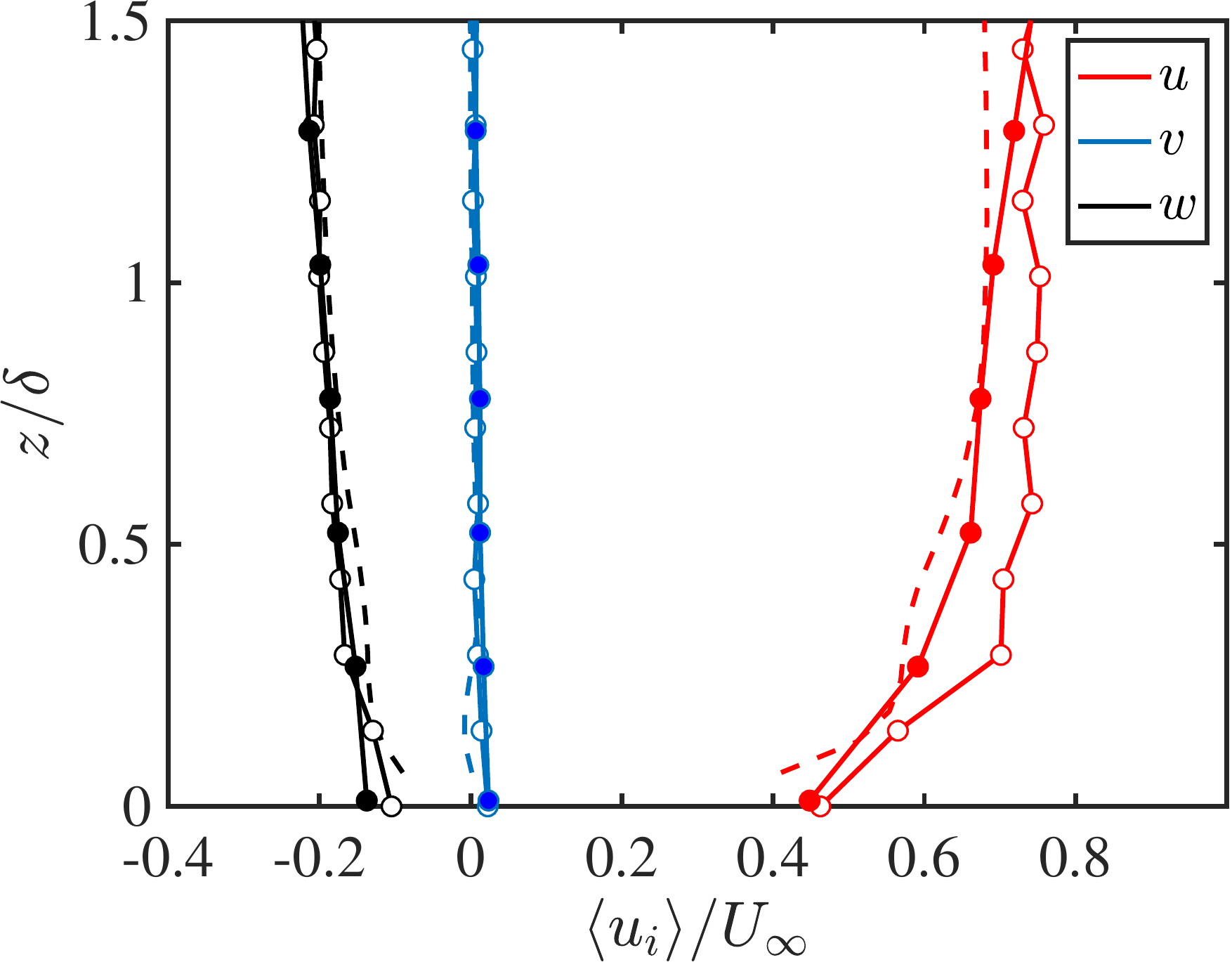}}
  \hspace{0.05cm}
  \subfloat[]{\includegraphics[width=0.32\textwidth]{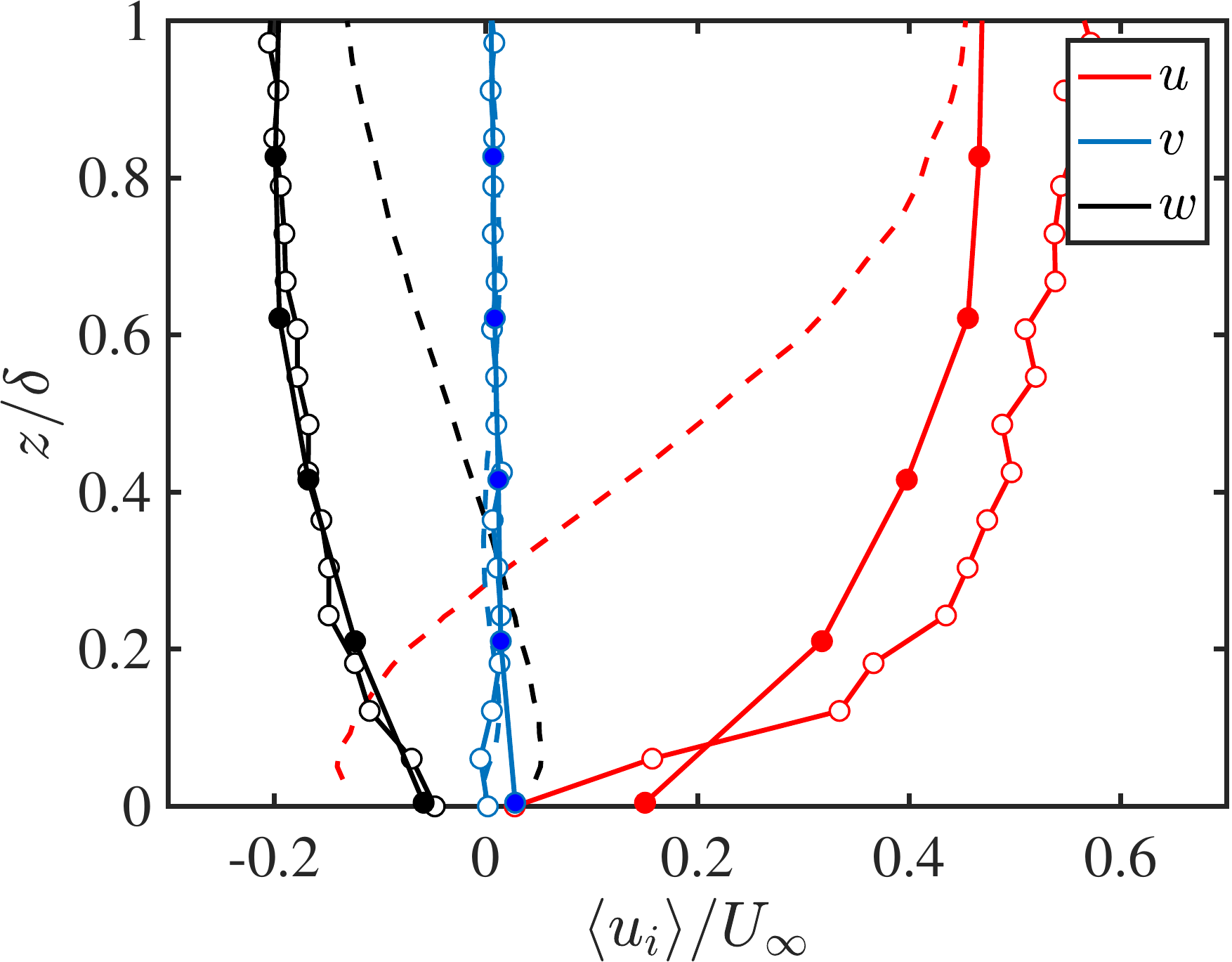}}
\end{center}
\caption{Mean velocity profiles for case C-N5-Rem8e3 (solid lines with
  $\bullet$) and C-D2 (solid lines with $\circ$) at (a) location
  1: upstream region of the fuselage $x=1168.4$~mm and $z=0$~mm (b)
  location 2 wing-body juncture $x=2747.6$~mm and $y=239.1$~mm, and (c)
  location 3: wing-body juncture close to the trailing edge at
  $x=2922.6$~mm and $y=239.1$~mm.  Experiments are denoted by dashed
  lines. Colors denote different velocity components. The distance $y$
  is normalized by the local boundary-layer thickness $\delta$ at that
  location.\label{fig:custom}}
\end{figure}

We evaluate potential improvements with BL-conforming
grids by comparing the results for case C-N5-Rem8e3 with case
C-D2. The grid for C-N5-Rem8e3 is generated for $N_{bl}=5$ and
$Re_\Delta^\mathrm{min}=10^4$ following the procedure described in
Section \ref{subsec:tbl}.  For reference, case C-D2 has 32 million
control volumes, whereas case C-N5-Rem8e3 has 11 million control
volumes. The mean velocity profiles for C-N5-Rem8e3 are shown in
Figure \ref{fig:custom} and compared with C-D2 at three
locations. Some moderate improvements are achieved at the fuselage
despite both cases sharing the same $N_{bl}$ at that location. The
improvements are accentuated at the juncture and trailing edge, where
C-N5-Rem8e3 outperforms C-D2 with less than a fourth of the grid
points per $\delta$ in each of the three spatial directions.

The results suggest that grids designed to specifically target the
boundary layer could improve the overall accuracy of WMLES. The
outcome might be explained by considering the upstream history of the
WMLES solution at a given station. Let's assume the simpler scenario
of WMLES of a flat-plate turbulent boundary layer along $x$ using two
grids: a constant-size grid and a BL-conforming grid. If we take an
$x$-location at which both grids have the same $N_{bl}$, the upstream
flow for the constant-size grid is underresolved compared to the
BL-conforming grid due to the thinner $\delta$ upstream the flow
(hence, less points per $\delta$ and larger errors). On the other
hand, the BL-conforming grid maintains a constant grid resolution
scaled in $\delta$ units and effectively more resolution upstream the
flow. Furthermore, even if at a given $x$ location $N_{bl}$ is larger
for the constant-size grid, the solution could be worse because of
error propagation from the upstream flow.  By assuming that
energy-containing eddies with lifetimes $\delta/U_\infty$ are advected
by $U_\infty$, we can estimate the downstream propagation of errors at
a given location $x_0$ by $\Delta x_e = \int_{x_0}^x \delta_0/\delta
\, \mathrm{d}x$, which is the streamwise distance required for the
energy-containing eddies to forget their past history. In ZPGTBL at
high $Re$~\citep{Sillero2013}, $\Delta x_e$ can reach values of
$\Delta x_e \approx 100\delta_0$. The long convective distance for
error propagation ($\sim 200$~mm for $\delta_0 \approx \Delta=2$~mm)
combined with the higher upstream errors for constant-size grids may
explain the improved results for C-N5-Rem8e3 reported in
Figure~\ref{fig:custom}.
%

\section{Conclusions}\label{sec:conclusions}

We have performed WMLES of the NASA Juncture Flow using charLES with
Voronoi grids. The simulations were conducted for an angle of attack
of $5^\circ$ and $Re=2.4 \times 10^6$. We have characterized the
errors in the prediction of mean velocity profiles and pressure
coefficient for three different locations over the aircraft: the
upstream region of the fuselage, the wing-body juncture, and the
wing-body juncture close to the trailing-edge. The last two locations
are characterized by strong mean-flow three-dimensionality and
separation.

The prediction of the pressure coefficient is below 5\% error for all
grid sizes considered, even when boundary layers were marginally
resolved.  We have shown that this good accuracy can be attributed to
the outer-layer nature of the mean pressure, which becomes less
sensitive to flow details within the turbulent boundary layer. A
summary of the errors incurred by WMLES in predicting mean velocity
profiles is shown in Figure \ref{fig:Errors_all} for the three
locations considered.  The message conveyed by the error analysis is
that WMLES performs as expected in regions where the flow resembles a
zero-pressure-gradient flat plate boundary layer. However, there is a
clear decline of the current models in the presence of wing-body
junctions and, more acutely, in separated zones. Moreover, the slow
convergence to the solution in these regions renders the brute-force
grid-refinement approach to improve the accuracy of the solution
unfeasible. The results reported above pertain to the mean velocity
profile predicted using the typical grid resolution for external
aerodynamics applications, i.e., 5--20 points per boundary-layer
thickness. The impact of the above deficiencies on the skin friction
prediction is uncertain due to the lack of experimental
measurements. Yet, it is expected that the errors observed in the mean
velocity profile would propagate to the wall stress.

Finally, we have shown that boundary-layer-conforming grids (i.e.,
grids maintaining a constant number of points per boundary-layer
thickness) allow for a more efficient distribution of grid points and
smaller errors. It was argued that the improved accuracy might be due
to the reduced propagation of WMLES errors in the streamwise
direction.  The increased accuracy provided by
boundary-layer-conforming grids will be investigated in more detail in
future studies along with the effect of targeted refinement in the
wakes. Our results suggest that novel modeling venues encompassing
physical insights, together with numerical and gridding advancements,
must be exploited to attain predictions within the tolerance required
for Certification by Analysis.

\section*{Acknowledgments}

A.L.-D. acknowledges the support of NASA under grant
No. NNX15AU93A. We thank Jane Bae and Konrad Goc for helpful comments.

\bibliographystyle{ctr}

\end{document}